\documentclass[useAMS,usenatbib]{mn2e}
\usepackage{times}
\usepackage{amsmath}
\usepackage{amsfonts}
\usepackage{rotating}
\usepackage{natbib}
\usepackage{amssymb}
\usepackage{subfig}
\usepackage{multirow}
\bibpunct{(}{)}{;}{a}{}{,}
\usepackage{animate}     
\usepackage{graphicx} 
\usepackage[toc,page]{appendix}     
                                                                   
\begin{document}
                                                                              
\title[The 43 GHz SiO masers on TX~Cam]{A long-term VLBA monitoring campaign of the 
\emission SiO masers toward TX~Cam\\ 
\begin{center} {\LARGE I. Morphology and Shock Waves}\end{center}} 
\author[Gonidakis et al.]{I. Gonidakis$^{1}$ \thanks{Present
contact: ioannis.gonidakis@csiro.au}, P. J. Diamond$^{2}$ \&
A. J. Kemball$^{3}$\\ 
$^{1}$CSIRO Astronomy and Space Science,
Vimiera and Pembroke Roads, Marsfield NSW 2122, Australia\\ 
$^{2}$SKA Organisation, 
Jodrell Bank Observatory, Lower Withington, Macclesfield, Cheshire, SK11 9DL, UK \\ 
$^{3}$Department of Astronomy,
University of Illinois at Urbana-Champaign,
1002 W. Green Street, Urbana, IL 61801, USA\\}

\newcommand{\etal}{et al.}
\newcommand{\kms}{km\,s$^{-1}$}
\newcommand{\HII}{H{\scriptsize II }}
\newcommand{\phase}{$\phi$} 
\newcommand{\similar}{$\sim$}
\newcommand{\emission}{{\it $v$=1, J=1$\rightarrow$0 }}
                                                                               
\pagerange{\pageref{firstpage}--\pageref{lastpage}} \pubyear{2011}
                                                                               
\maketitle
                                                                               
\label{firstpage}
                                                                               
\begin{abstract}

We present the latest and final version of the movie of the SiO masers toward the
Mira variable TX~Cam. The new version consists of 112 frames (78 successfully reduced 
epochs) with data covering almost three complete stellar cycles between 24$^{th}$ May 
1997 and 25$^{th}$ January 2002, observed with the VLBA. In this paper we examine the 
global morphology, kinematics and variability of the masering zone. The morphology of 
the emission is confined in a structure that usually resembles a ring or an ellipse, 
with occasional deviations due to localised phenomena. The ring appears to be 
contracting and expanding, although for the first cycle contraction is not observed.
The width and outer boundary of the masering zone follow the stellar pulsation. 
Our data seem to be consistent with a shock being created once per stellar cycle at 
maximum that propagates with a velocity of $\sim$7 \kms. The difference in velocities 
along different axes strongly suggests that the outflow in TX Cam is bipolar. The 
contribution of projection is examined and our results are compared with the latest 
theoretical model. 

\end{abstract}
                                                                               
\begin{keywords}
stars: variables: other -- stars: AGB and post-AGB -- stars: circumstellar matter -- 
masers -- shock waves -- radio lines: stars
\end{keywords}
                                                                               
\section{Introduction}
The discovery of pulsating variability in stars can be traced back to 1596, when David 
Fabricius  \citep{wolf_1} first observed the phenomenon in an individual star that later 
would give its name to an entire variability class. Johann Fokkens Holwarda in his book 
Historiola Mirae Stellae named the star Mira, the latin world for Wonderful 
\citep{poggendorff_1}. We now know that these ``Wonderful'' Variables are evolved red 
giants, located on the Asymtotic Giant Branch (AGB) of the H-R diagram. Mira Variables 
mark a stage in the evolutionary path of stars with zero age main sequence masses 
between of 1$<M_{\sun}<$8 that pulsate with regular periodicity and have fluctuation
in the V-band greater than 2.5$^{m}$ \citep{habing_1,habing_2}. Regarding the structure 
of typical Miras, they have a radius of 2 AU or more and they consist of a degenerated 
C/O core, surrounded by a thin layer of helium, that is in turn covered by a huge 
hydrogen mantle \citep{habing_1}. Due to their high mass-loss rates (10$^{-7}$ to 
10$^{-4}$ M$_{\sun}$) they are surrounded by an extended Circumstellar Envelope (CSE) 
\citep{olofsson_1}. The near-circumstellar morphology is not necessarily symmetric 
however \citep{josselin_1, thompson_1}. Observations with the Infrared Optical telescope 
Array by \cite{ragland_1} show that one third of the Miras in their sample were found to 
have significant asymmetries.

The total luminosity of Miras vary by only approximately an order of magnitude; 
However, in the visual part of the spectrum they can vary by up to 8 orders of 
magnitude. Their relatively low photospheric surface temperatures (2000-3500 K) allow 
the formation of molecules in their CSEs during minima, including metalic oxides which 
absorb wide bands of the optical spectrum, resulting in the extreme visual variabilty 
observed \citep{reid_1}.
 
SiO masers are located between the dust formation region and the stellar atmosphere 
\citep{elitzur_2}, in a very complex area known as the extended atmosphere. 
\cite{wittkowski_1} conducted the first multi-epoch mid-infrared and radio 
interferometry study of a Mira star (S Ori), in order to determine the photospheric 
radius, the dust shell parameters and the SiO maser spot location and examine 
dependences on the pulsation phase. They concluded that, during minimum visual phase, 
dust (consisting only of Al$_{2}$O$_{3}$) and SiO masers are co-located, forming close 
to the surface. However, after the visual maximum, the inner dust shell is expanded 
by $\sim 20\%$ while the maser spots remain at about the same location.  

The first high resolution images of the 43 GHz SiO masers around late-type stars were 
produced by \cite{miyoshi_1}, \cite{diamond_2} and \cite{greenhill_1} and revealed that 
the masers are confined to well defined projected rings, overruling the prevailing belief 
at that time that SiO masers form chaotic structures around variable stars. The VLBA 
observations of the Miras TX Cam and U Her by \cite{diamond_2} showed that SiO masers 
form at 2-4 R$_{\star}$, placing them in the extended atmosphere of the stars, inside the 
dust formation zone. The main kinematic behaviour appeared to be that of outflow. 
The emission pattern was consistent with tangential amplification. Subsequent to these 
initial reported observations, a synoptic monitoring campaign was commenced for a 
limited number of sources.

The first movie of the TX Cam monitoring campaign was published by \cite{diamond_1}. 
This paper included the first 41 epochs as a movie of 44 frames covering a complete 
stellar pulsation cycle at an angular resolution of $\sim$0.1 mas and revealed the gross 
kinematic properties of the SiO maser emission. The morphology and properties of the 
shell varied with time and stellar phase. Individual maser features persisted over many 
epochs and the predominant kinematic behaviour of the ring was that of expansion. 
Contrary to the models that assume spherical symmetry, the structure and evolution of 
the ring revealed a high degree of asymmetry. There was also evidence of ballistic 
deceleration and proper motion analysis revealed motions between $\sim$5-10 \kms, 
distributed around the ring.

An expanded 73-frame movie (60 epochs) by \cite{gonidakis_1} uncovered more properties 
of the morphology and the kinematics of the masering shell. Over two pulsation cycles, 
the movie revealed that contraction occurred during the second cycle. The time-span of 
the movie allowed the study of short- and long-term variability properties, i.e. changes 
not only within a cycle but also from one cycle to another. The 43 GHz flux variability 
was found to follow that of the optical with a $\sim$10\% lag but the flux densities were 
uncorrelated. The width of the ring was also found to be correlated with the stellar 
pulsation phase but there was no correlation found between the velocities and position 
angle of the ring. The lifetime of individual components were found to follow a Gaussian 
distribution with a peak between 150 and 200 days, while the spectra were dominated by 
blue and red-shifted peaks that formed at different times in the stellar cycle and had 
different lifetimes. 

The paper is organised as follows. Section 2 outlines the observations and data 
reduction. In Section 3 the new version of the movie is presented, as well as the 
analysis of the data and the results. Section 4 examines the contribution of shock waves 
to the overall structure and kinematics. The science results are discussed in Section 5 
and the paper ends with the conclusions of this study in Section 6.

\section{Observations and data reduction}
The monitoring of the \emission masers on TX Cam started on the 24$^{th}$ of May 1997 with
the Very Long Baseline Array (VLBA\footnote{The Very Long Baseline Array (VLBA) is 
operated by the National RadioAstronomy Observatory (NRAO), a facility of the National 
Science Foundation (NSF) under cooperative agreement by Associated Universities, Inc.}), 
and one antenna from the VLA. Until the 9$^{th}$ of September 1999,  observations were 
conducted in biweekly intervals. From that date until the end of the project on the 
25$^{th}$ of January 2002, the intervals were monthly. Given a pulsation period of 557 
days for TX Cam \citep{kholopov_1} they correspond to a coverage of 3.06 pulsation 
periods. The result was 82 individual datasets, 78 of which were reduced successfully. 
Due to bad weather conditions over parts of the array or loss of crucial antennae near 
the array centre, some epochs did not produce images of acceptable quality and had to 
be discarded. 
 
All epochs were observed at a rest frequency of 43.122027 GHz, centred at an LSR 
velocity of 9.0 \kms. Each observing run was 6-8 hours long, with 2.5-3.5 hours devoted 
to TX Cam and the rest to calibrators and other target sources. Data were recorded in 
dual polarisation over a 4 MHz bandwidth and correlated with the VLBA correlator in 
Socorro, creating datasets in all 4 cross-polarisation pairs over 128 spectral channels. 
The nominal spectral resolution was 31.25 kHz, corresponding to a velocity resolution 
of 0.217 \kms.

The science goals of the project required consistency and uniformity in both 
observations and data analysis. A uniform observing strategy was assured by retaining 
the same configuration during each experiment as described in the previous paragraph. 
The requirement for consistent, uniform data analysis demanded the creation of
an automated data reduction procedure. This was developed as a POPS script within the 
AIPS package and was based on the techniques described by \cite{kemball_1} and 
\cite{kemball_2}. The pipeline was divided into logical steps that demanded a minimum 
of user interaction; this was limited to the examination of the results at the end of 
each step for quality assurance and the interactive data editing. The final pipeline 
result for each epoch was a 128-channel cube of 1024$\times$1024 pixels in Stokes 
\{I,Q,U,V\} on the projected plane of the sky with a pixel separation of 0.1 mas.

\section{The SiO Masers}

	\subsection{The Movie}

	One of the objectives of this project was to create a movie to serve as a realistic 
	visualisation of the evolution and distribution of the SiO masers around TX Cam. 
	The final version of the movie consists of 112 frames and covers 3.06 stellar 
	cycles. It would be useful to distinguish between \emph{epochs} and \emph{frames}, 
	since these terms will be frequently used in the following paragraphs. The term 
	\emph{epoch} corresponds to maps produced by actual observations while the term 
	\emph{frame} includes also the interpolated maps. Thus, the whole project consists 
	of 82 epochs, 78 of which were successfully reduced, 34 interpolated frames due to 
	bad or missing data, resulting into a 112-frame movie.
        
		\subsubsection{Post-production}
		As discussed in the previous section, the data were recorded and reduced 
		identically in order to ensure homogeneity. Further, uniform image 
		post-processing is required in order to render a successful movie visualisation 
		of the maser morphological evolution.The final cubes were averaged over 
		frequency using task {\scriptsize SQASH} in AIPS giving the final total 
		intensity image for each epoch. In order to suppress spurious features in the 
		noise floor and in order to increase contrast, a cutoff level was applied with 
		task {\scriptsize BLANK} in AIPS and each frame was converted into its square 
		root with task {\scriptsize MATHS}. This conversion was only applied to the 
		frames used for the compilation of the movie; for the data analysis the total 
		intensity maps were used. The extent of the maser emission from the 
		circumstellar shell occupied only a sub-region within each 1024$\times$1024 
		pixel frame, so the final image processing step was to trim them down to 
		680$\times$680 pixels using task {\scriptsize SUBIM}.  
		
		\begin{table*}

		\caption{List of epochs, observation dates and the corresponding phase
		of the star. Comments in the status column indicate missing or failed
		epochs. These epochs were interpolated in order to have a smooth, time
		coherent movie. The time when observations switched to a monthly
		interval is indicated in the table.}

		\center
		\begin{tabular}{cccc||cccc}

		\hline
		\hline

		\bf Epoch code   & \bf Observing date           & \bf Optical phase   & \bf Status & \bf Epoch code   & \bf Observing date           & \bf Optical phase   & \bf Status\\
		\hline
		
		\multicolumn{4}{c}{Observations every 2 weeks}\ & \multicolumn{4}{c}{Observations every month}  \\

		\hline
		BD46A            & 1997  May  24                & 0.68 $\pm$ 0.01     & $\surd$ & BD62A            & 1999 September 9             & 2.18 $\pm$ 0.01     & $\surd$\\
		BD46B            & 1997  June  7                & 0.70 $\pm$ 0.01     & $\surd$  & BD62AB           & 1999 September 27            & 2.21 $\pm$ 0.01     & $\between$ \\
		BD46C            & 1997  June  22               & 0.73 $\pm$ 0.01     & $\surd$ & BD62B            & 1999 October 15              & 2.24 $\pm$ 0.01     & $\times$\\
		BD46D            & 1997  July  6                & 0.75 $\pm$ 0.01     & $\surd$ & BD62BC           & 1999 October 31              & 2.27 $\pm$ 0.01     & $\between$ \\
		BD46E            & 1997  June 19                & 0.78 $\pm$ 0.01     & $\surd$ & BD62C            & 1999 November 14             & 2.30 $\pm$ 0.01     & $\surd$\\
		BD46F            & 1997  August  2              & 0.80 $\pm$ 0.01     & $\surd$ & BD62CD           & 1999 December 2              & 2.33 $\pm$ 0.01     & $\between$ \\
		BD46G            & 1997  August  16             & 0.83 $\pm$ 0.01     & $\surd$ & BD62D            & 1999 December 19             & 2.36 $\pm$ 0.01     & $\surd$\\
		BD46H            & 1997  August  28             & 0.85 $\pm$ 0.01     & $\surd$ & BD62DE           & 2000 January 1               & 2.38 $\pm$ 0.01     & $\between$\\
		BD46I             & 1997  September  12          & 0.87 $\pm$ 0.01     & $\surd$ & BD62E            & 2000 January 15              & 2.41 $\pm$ 0.01     & $\surd$\\
		BD46J            & 1997  September  26          & 0.90 $\pm$ 0.01     & $\surd$ & BD62EF           & 2000 January 31              & 2.44 $\pm$ 0.01     & $\between$ \\
		BD46K            & 1997  October  9             & 0.92 $\pm$ 0.01     & $\surd$ & BD62F            & 2000 February 14             & 2.46 $\pm$ 0.01     & $\surd$\\
		BD46L            & 1997  October  25            & 0.95 $\pm$ 0.01     & $\times$ &  BD62FG           & 2000 March 2                 & 2.49 $\pm$ 0.01     & $\between$ \\
		BD46M            & 1997  November  8            & 0.98 $\pm$ 0.01     & $\surd$ & BD62G            & 2000 March 17                & 2.52 $\pm$ 0.01     & $\surd$\\
		BD46N            & 1997  November 21            & 1.00 $\pm$ 0.01     & $\surd$ & BD62GH           & 2000 April 4                 & 2.55 $\pm$ 0.01     & $\between$ \\
		BD46O            & 1997  December  5            & 1.03 $\pm$ 0.01     & $\surd$ & BD62H            & 2000 April 21                & 2.58 $\pm$ 0.01     & $\surd$\\
		BD46P            & 1997  December  17           & 1.05 $\pm$ 0.01     & $\surd$ & BD62HI           & 2000 May 5                   & 2.61 $\pm$ 0.01     & $\between$ \\
		BD46Q            & 1997  December  30           & 1.07 $\pm$ 0.01     & $\times$ & BD62I            & 2000 May 21                  & 2.64 $\pm$ 0.01     & $\surd$\\
		BD46R            & 1998  January  13            & 1.10 $\pm$ 0.01     & $\surd$ & BD62IA           & 2000 June 6                  & 2.67 $\pm$ 0.01     & $\between$\\
		BD46S            & 1998  January  25            & 1.12 $\pm$ 0.01     & $\surd$ & BD69A            & 2000 June 22             & 2.70 $\pm$ 0.01     & $\surd$\\
		BD46T            & 1998  February  5            & 1.14 $\pm$ 0.01     & $\surd$ & BD69AB           & 2000 July 4            & 2.72 $\pm$ 0.01     & $\between$ \\
		BD46U            & 1998  February  22           & 1.17 $\pm$ 0.01     & $\surd$ & BD69B            & 2000 July 16              & 2.74 $\pm$ 0.01     & $\surd$\\
		BD46V            & 1998  March  5               & 1.19 $\pm$ 0.01     & $\surd$ & BD69BC           & 2000 August 3              & 2.77 $\pm$ 0.01     & $\between$ \\
		BD46VW        & 1998  March  21              & 1.22 $\pm$ 0.01     & $\between$ & BD69C            & 2000 August 21             & 2.81 $\pm$ 0.01     & $\times$\\
		BD46W           & 1998  April  7               & 1.24 $\pm$ 0.01     & $\surd$ & BD69CD           & 2000 September 6              & 2.84 $\pm$ 0.01     & $\between$ \\
		BD46X            & 1998  April  19              & 1.27 $\pm$ 0.01     & $\surd$ & BD69D            & 2000 September 21             & 2.86 $\pm$ 0.01     & $\surd$\\
		BD46Z            & 1998  May  10                & 1.30 $\pm$ 0.01     & $\surd$  & BD69DE           & 2000 October 5               & 2.89 $\pm$ 0.01     & $\between$\\
		BD46AA         & 1998  May  22                & 1.33 $\pm$ 0.01     & $\surd$ & BD69E            & 2000  October 19            & 2.91 $\pm$ 0.01     & $\surd$\\
		BD46AB         & 1998  June 6                 & 1.35 $\pm$ 0.01     & $\surd$ & BD69EF           & 2000 November 2              & 2.94 $\pm$ 0.01     & $\between$ \\
		BD46AC         & 1998  June  18               & 1.37 $\pm$ 0.01     & $\surd$ & BD69F            & 2000 November 17            & 2.96 $\pm$ 0.01     & $\surd$\\
		BD46AD         & 1998  July  3                & 1.40 $\pm$ 0.01     & $\surd$  & BD69FG           & 2000 December 3                 & 2.99 $\pm$ 0.01     & $\between$ \\
		BD46AE         & 1998  July 23                & 1.44 $\pm$ 0.01     & $\surd$ & BD69G            & 2000 December 20                & 3.02 $\pm$ 0.01     & $\surd$\\
		BD46AF         & 1998  August  9              & 1.47 $\pm$ 0.01     & $\surd$ & BD69GH           & 2001 January 1                 & 3.05 $\pm$ 0.01     & $\between$ \\
		BD46AG        & 1998  August  23             & 1.49 $\pm$ 0.01     & $\surd$ & BD69H            & 2001 January 20                & 3.08 $\pm$ 0.01     & $\surd$\\
		BD46AH        & 1998  September  9           & 1.52 $\pm$ 0.01     & $\surd$ & BD69HI           & 2001 February 3                   & 3.10 $\pm$ 0.01     & $\between$ \\
		BD46AI          & 1998  September  25          & 1.55 $\pm$ 0.01     & $\surd$ & BD69I            & 2001 February 18                  & 3.13 $\pm$ 0.01     & $\surd$\\
		BD46AJ         & 1998  October  14            & 1.59 $\pm$ 0.01     & $\surd$ & BD69IJ           & 2001 March 3              & 3.15 $\pm$ 0.01     & $\between$ \\
		BD46AK         & 1998  October  29            & 1.61 $\pm$ 0.01     & $\surd$ & BD69J            & 2001 March 16             & 3.18 $\pm$ 0.01     & $\surd$\\
		BD46AL         & 1998  November 17            & 1.65 $\pm$ 0.01     & $\surd$ & BD69JK           & 2001 March 31               & 3.20 $\pm$ 0.01     & $\between$\\
		BD46AM        & 1998  December  6            & 1.68 $\pm$ 0.01     & $\surd$ & BD69K            & 2001 April 15              & 3.23 $\pm$ 0.01     & $\surd$\\
		BD46AN        & 1998  December  23           & 1.71 $\pm$ 0.01     & $\surd$ & BD69KL           & 2001 May 2               & 3.26 $\pm$ 0.01     & $\between$ \\
		BD46AO        & 1999  January  5             & 1.74 $\pm$ 0.01     & $\surd$ & BD69L            & 2001 May 19             & 3.29 $\pm$ 0.01     & $\surd$\\
		BD46AP        & 1999  January  23            & 1.77 $\pm$ 0.01     & $\surd$ & BD69LM           & 2001 June 6                 & 3.32 $\pm$ 0.01     & $\between$ \\
		BD46AQ        & 1999  February  6            & 1.79 $\pm$ 0.01     & $\surd$ & BD69M            & 2001 June 24                & 3.36 $\pm$ 0.01     & $\surd$\\
		BD46AR         & 1999  February  19           & 1.82 $\pm$ 0.01     & $\surd$ & BD69MMN           & 2001 July 7                 & 3.38 $\pm$ 0.01     & $\between$ \\
		BD57A            & 1999 March 12                & 1.85 $\pm$ 0.01     & $\surd$ & BD69MN           & 2001 July 20                 & 3.40 $\pm$ 0.01     & $\between$ \\
		BD57B            & 1999 March 29                & 1.88 $\pm$ 0.01     & $\times$ & BD69MNN           & 2001 August 3                 & 3.43 $\pm$ 0.01     & $\between$ \\
		BD57C            & 1999 April 12                & 1.91 $\pm$ 0.01     & $\surd$ & BD69N            & 2001 August 16                & 3.45 $\pm$ 0.01     & $\surd$\\
		BD57D            & 1999 April 24                & 1.93 $\pm$ 0.01     & $\surd$ & BD69NA           & 2001 September 7                 & 3.49 $\pm$ 0.01     & $\between$ \\
		BD57E            & 1999 May 14                  & 1.97 $\pm$ 0.01     & $\surd$ & BD79A            & 2001 September 28            & 3.53 $\pm$ 0.01     & $\surd$\\
		BD57F            & 1999 May 31                  & 2.00 $\pm$ 0.01     & $\surd$ & BD79B            & 2001 October 12              & 3.55 $\pm$ 0.01     & $\surd$\\
		BD57G            & 1999 June 12                 & 2.02 $\pm$ 0.01     & $\surd$ & BD79BC           & 2001 October 27              & 3.58 $\pm$ 0.01     & $\between$ \\
		BD57H            & 1999 June 27                 & 2.05 $\pm$ 0.01     & $\surd$ & BD79C            & 2001 November 12             & 3.61 $\pm$ 0.01     & $\surd$\\
		BD57I              & 1999 July 12                 & 2.07 $\pm$ 0.01     & $\surd$ & BD79CD           & 2001 December 1              & 3.64 $\pm$ 0.01     & $\between$ \\
		BD57J            & 1999 July 30                 & 2.11 $\pm$ 0.01     & $\surd$ & BD79D            & 2001 December 18             & 3.67 $\pm$ 0.01     & $\surd$\\
		BD57K            & 1999 August 11               & 2.13 $\pm$ 0.01     & $\surd$ & BD79DE           & 2002 January 6               & 3.71 $\pm$ 0.01     & $\between$\\
		BD57L            & 1999 August 24               & 2.15 $\pm$ 0.01     & $\surd$ & BD79E            & 2002 January 25              & 3.74 $\pm$ 0.01     & $\surd$\\
		
		\hline
		\hline
		\end{tabular}
		\label{tab_1}
	
		\medskip
		\vspace{-0.2cm}
		\flushleft{$\surd$ indicates successfully reduced epochs}
		\vspace{-0.2cm}
		\flushleft{$\times$ indicates interpolated epochs due to bad data}
		\vspace{-0.2cm}
		\flushleft{~$\between$ indicates interpolated epochs; observations were not 
		scheduled at these times.}

		\end{table*}

		\begin{figure*}
		\centering
		\subfloat[]{\label{fig:surface}\includegraphics[trim=0mm 0mm 0mm 40mm, clip, scale=0.45]{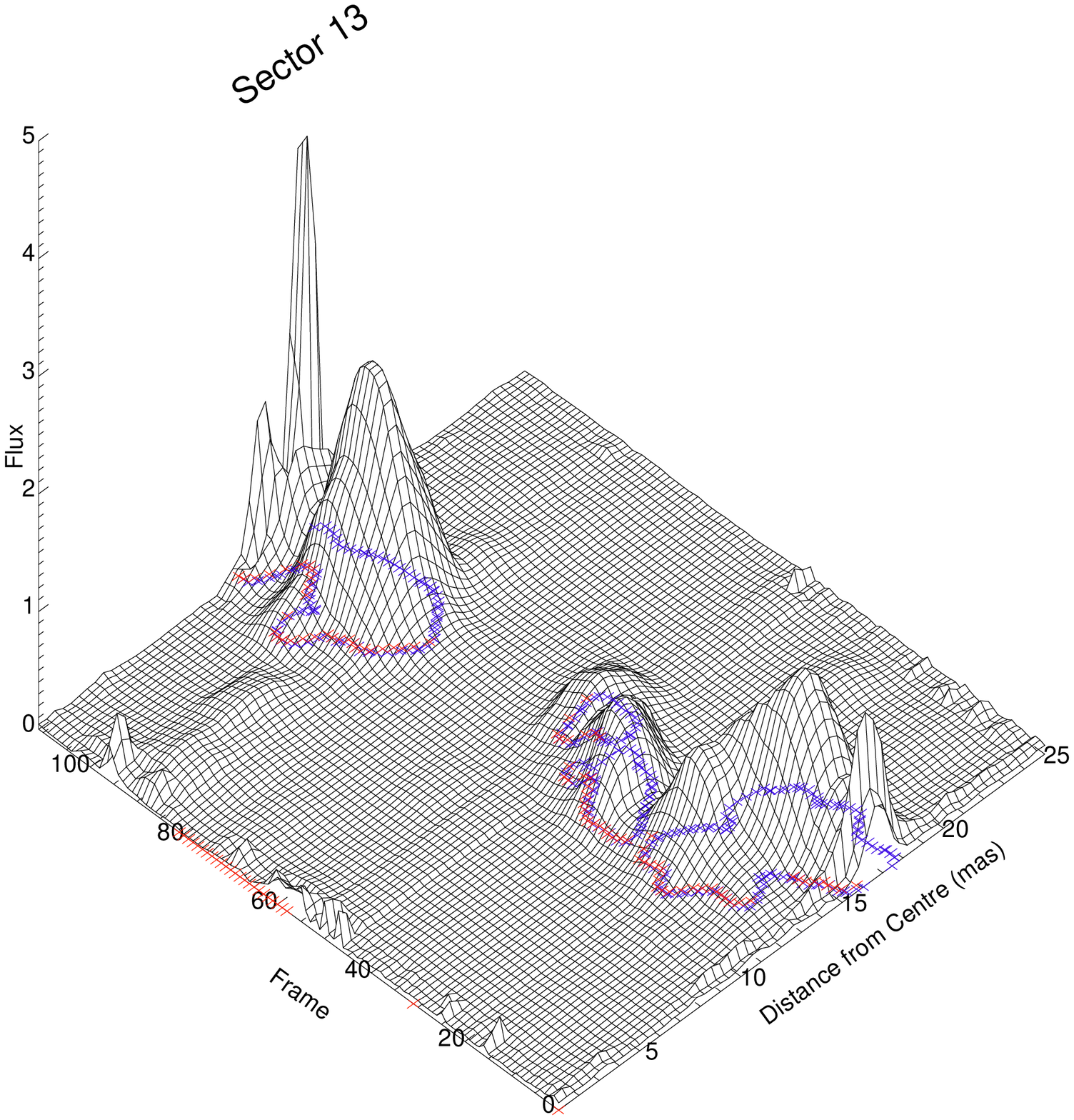}}
		\subfloat[]{\label{fig:ring}\includegraphics[trim=0mm 0mm 0mm 7mm, clip, scale=0.45]{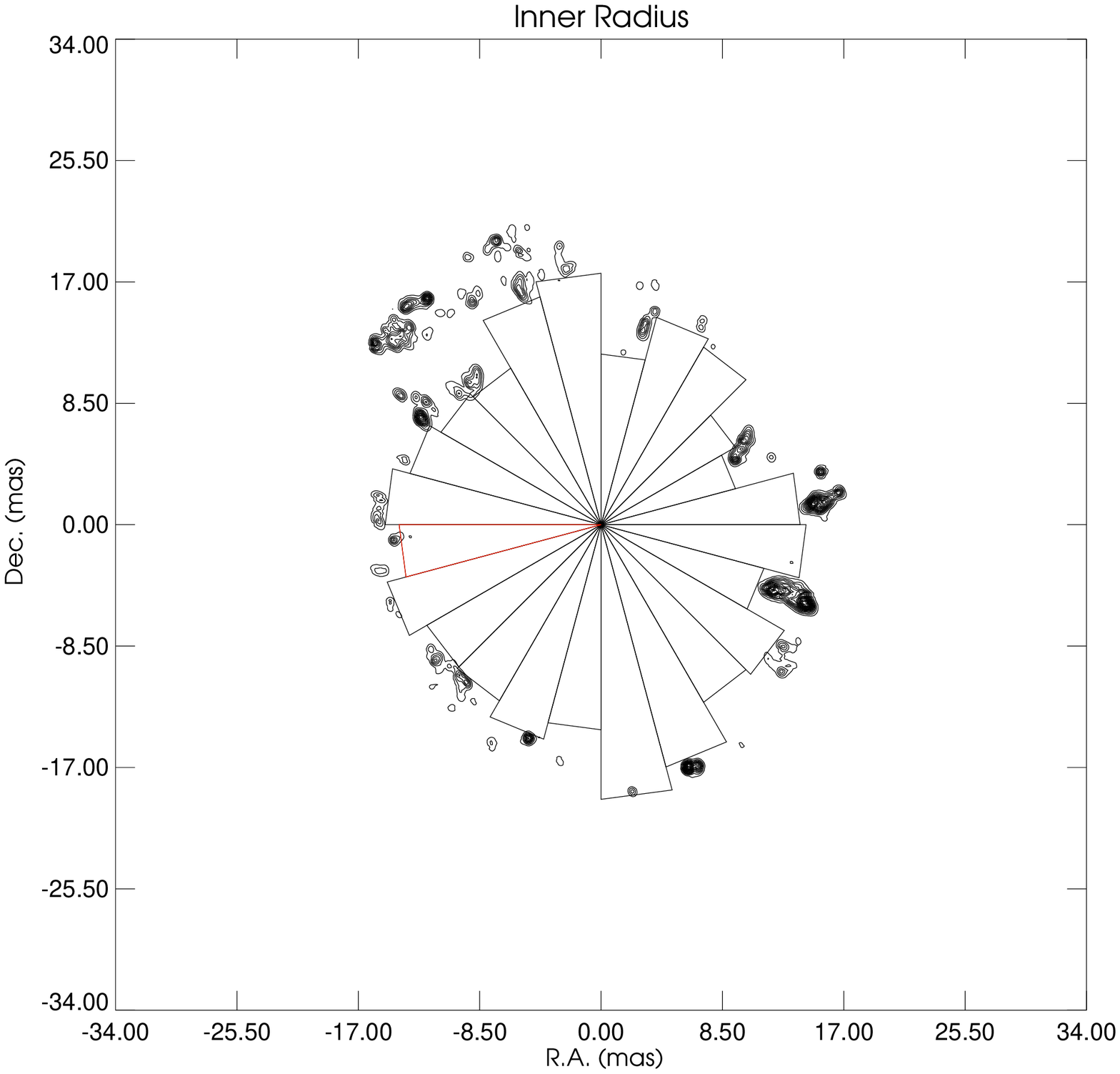}}
		\caption{(a) An example of the technique used for the determination of the inner 
		shell boundary for Sector 13. The surface plot is the concatenation of all the 
		radial intensity profiles over time. The blue contours trace the point with 
		brightness three time that of the off-source regions and the red superimposed 
		crosses correspond to the estimated inner boundary. (b) Total intensity image 
		for frame 55 with the 24 sectors over-plotted as triangles with their apex at 
		the centre of the image and height equal to the calculated radius for each 
		sector. Sector 13 is plotted with red color.}
		\label{fig:fig_1}
		\end{figure*}

		However, the above image processing steps leave two additional unsolved 
		problems, namely the lack of spatial registration between successive frames and 
		non-uniform time sampling over the time series of epochs. The first issue arises 
		from the fact that any information on the absolute position is lost during 
		self-calibration. The second is caused by the switch from biweekly to monthly 
		observing intervals and arises also from epochs with missing data, creating gaps 
		in the movie. These spatial and temporal alignment issues were overcome by 
		developing appropriate software. Each pair of consecutive epochs was first 
		superimposed by eye to establish an approximate shift that should be applied to 
		the second image, in order to register it with the first. After this initial 
		alignment, sub-pixel corrections were calculated using task {\scriptsize LGEOM} 
		in AIPS (adapted from a program written by Craig Walker). All the corrections 
		were applied using task {\scriptsize LGEOM} in AIPS. A detailed explanation of 
		the alignment technique can be found at \cite{diamond_1}.
		
		After the completion of the spatial alignment, the final image processing step 
		was to populate the movie with the missing frames and interpolate it, where 
		needed, onto a regularly spaced set of time intervals. For this purpose we used 
		task {\scriptsize COMB} in AIPS, to interpolate missing frames as the average of 
		adjacent frames. The frames were concatenated into movie files (in .mov video 
		format) with \emph{ffmpeg}, a free program for handling multimedia data. Table 
		\ref{tab_1} lists all the frames used to the compilation of this movie into two 
		sets of four columns. In each set, the first column provides 
		the code name for each epoch, the second the date of observation, the third the 
		corresponding phase, calculated from the optical maximum provided by \cite{gray_1} 
		and the fourth indicates whether the epoch was observed and if it was reduced 
		successfully. The result was a 112-frame movie, covering 3.06 pulsation cycles of 
		TX Cam, the first frame of which can be viewed at Fig.~\ref{fig_movie_imax}. 
		The movie is provided as online supplementary material of this paper.
 
		\subsubsection{The New Release}
		{\bf Structure:} As reported in the previous TX Cam studies published 
		(\cite{diamond_2}, \cite{diamond_1} and \cite{gonidakis_1}) the emission is 
		confined in a projected ring-like structure. Although predominantly circular, 
		the ring can change its shape with time, resembling at times an ellipse, an arc, 
		or even a complicated figure-eight structure. The width of the ring changes with 
		time as does the radius at which it is located relative to its host star.  There 
		are three distinct ways for the emission to manifest itself; in elongated {\it 
		filaments}, in smaller and more circular {\it spots} and in faint {\it diffuse} 
		emission. The first two are most common in the inner portion of the ring while 
		diffuse emission is dominant in its outermost parts, defining an abstract outer 
		boundary.  The lack of emission within the ring indicates that the amplification 
		of the masers is tangential \citep{diamond_2}. \\
		{\bf Kinematics:} The projected ring radius changes with time and significant 
		individual proper motions are apparent. Generally, the flow of the material 
		appears ordered, mainly revealing expansion or contraction, there are though 
		localised deviations from this behaviour. A closer examination reveals mainly four 
		discrete apparent kinematic motions, {\it outflows}, {\it inflows}, {\it 
		ricochets} (features that bounce and change their trajectories, contrary to the 
		ordered kinematic behaviour of their surroundings), and {\it splits} (parts of 
		the ring that split into two new arcs, one following the overall motion of the 
		ring and the other moving to the opposite direction). The first two are global, 
		lasting for long fractions of the pulsation period. The latter are rare and 
		localised but they are major contributors in forming the overall shape of the 
		emission. 

		\begin{figure}
		\centering
		\subfloat[]{\label{fig:radius}\includegraphics[trim=8mm 0mm 0mm 0mm, clip, scale=0.45]{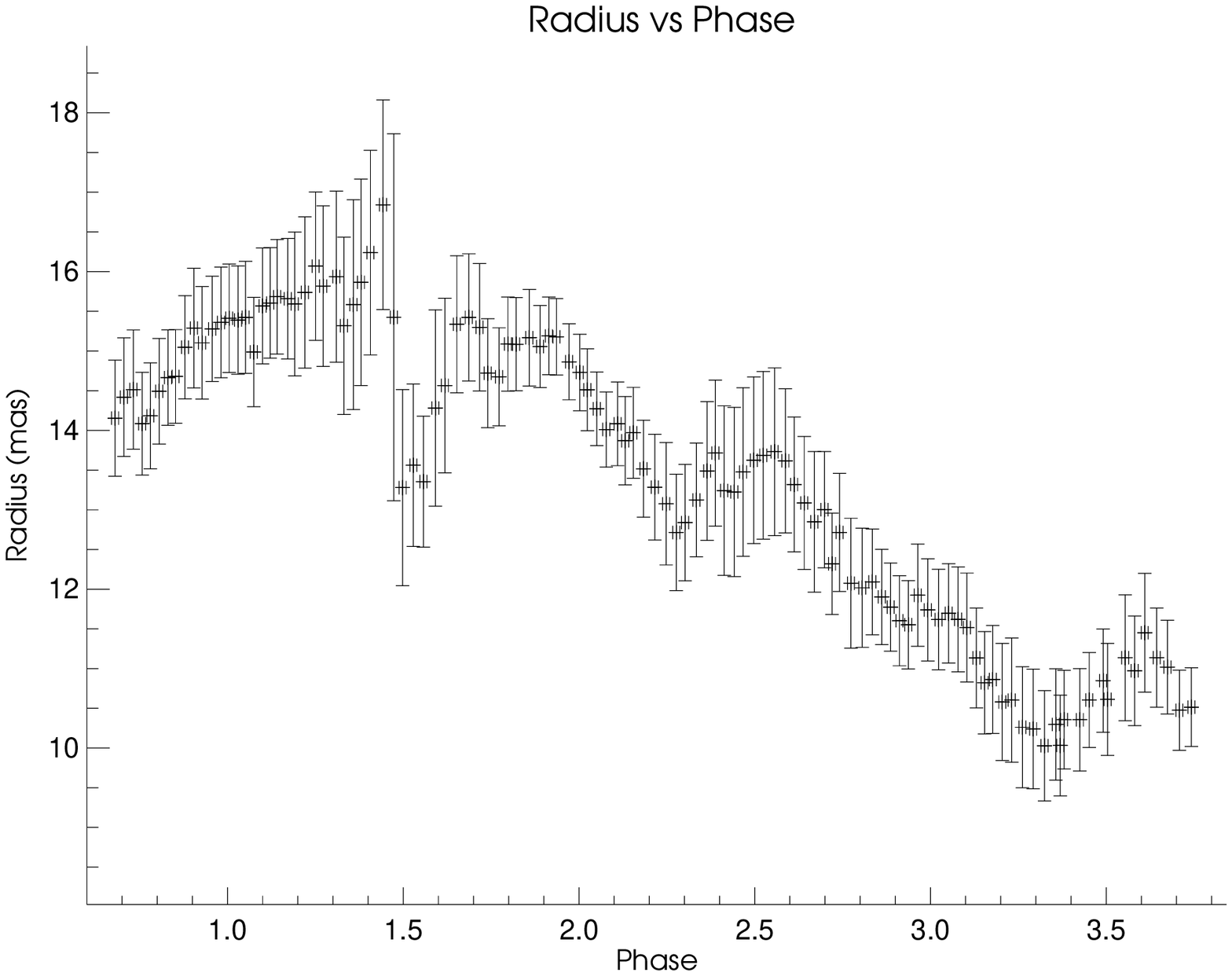}}\\
		\subfloat[]{\label{fig:distance}\includegraphics[trim=8mm 0mm 0mm 0mm, clip, scale=0.45]{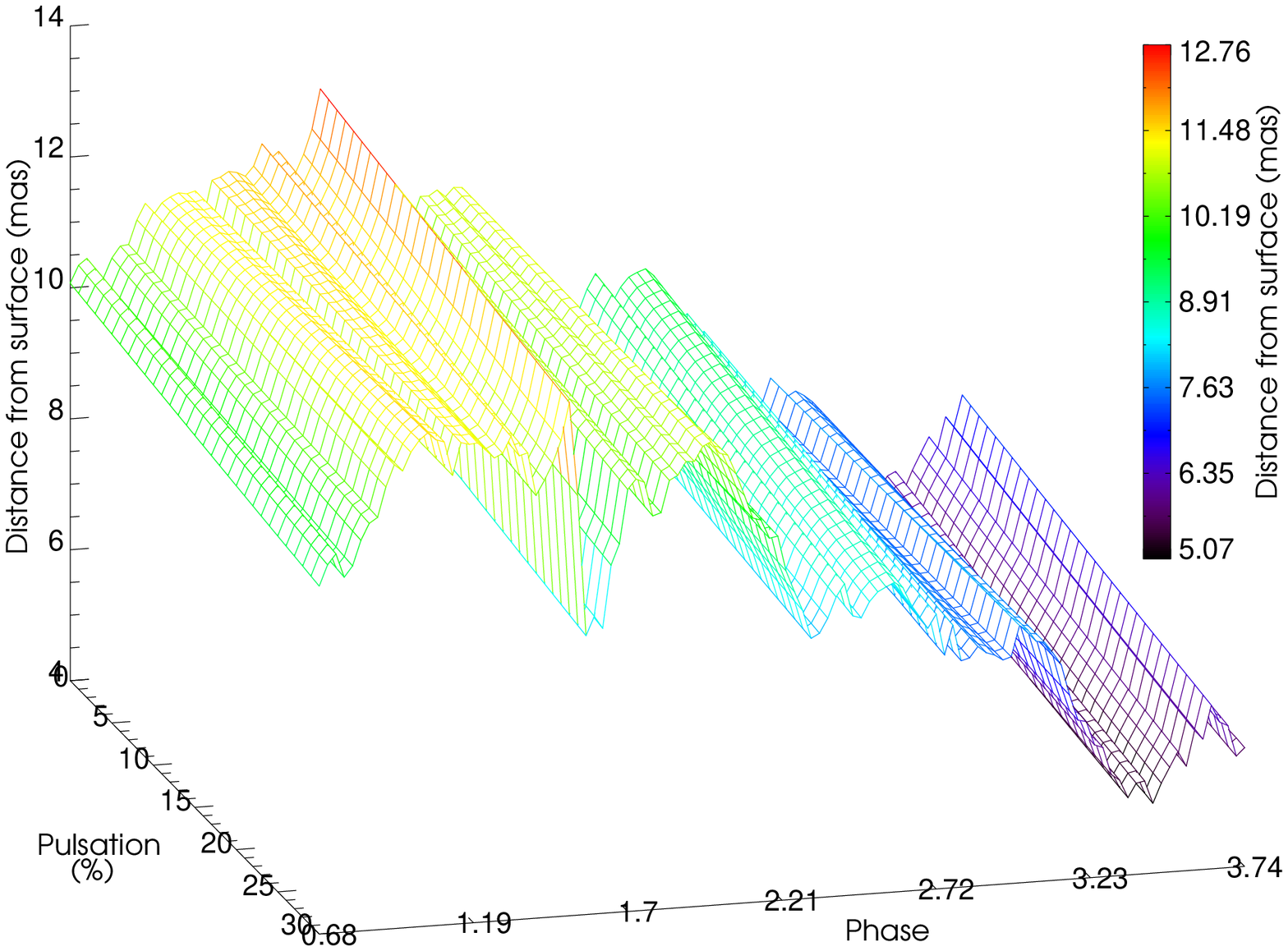}}\\
		\subfloat[]{\label{fig:rad_vs_int}\includegraphics[trim=8mm 0mm 0mm 0mm, clip, scale=0.45]{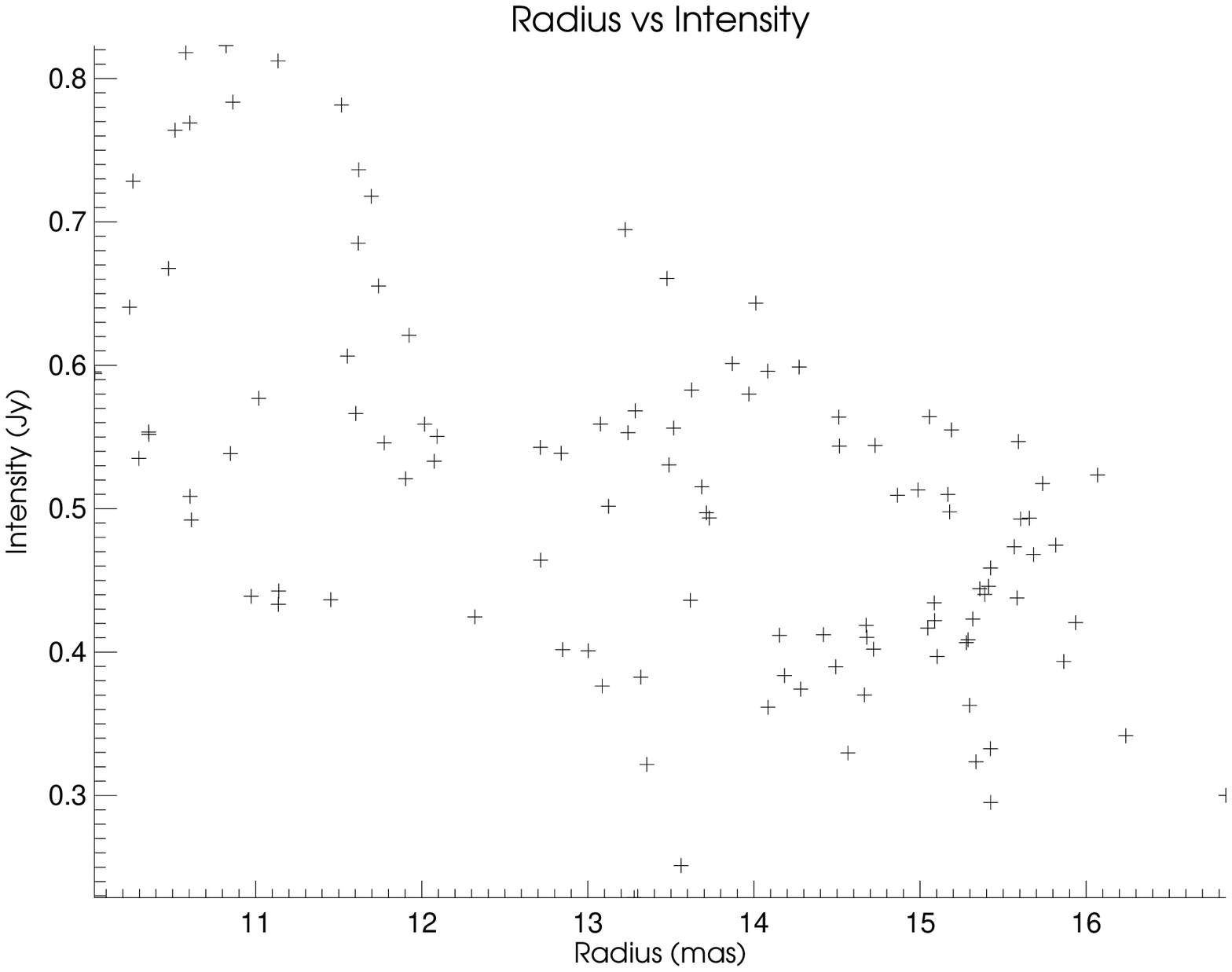}}\\
		\caption{(a) The inner shell radius plotted as a function of stellar pulsation. 
		(b) The distance of the inner boundary from the surface of the star for 
		different pulsation amplitudes (see text for adopted stellar radius and pulsation 
		profile). (c) The sector-averaged intensity of the ring plotted with radius.}
		\label{fig:fig_2}
		\end{figure}
		\begin{figure}
		\centering
		\subfloat[]{\label{fig:width}\includegraphics[trim=8mm 0mm 0mm 0mm, clip, scale=0.45]{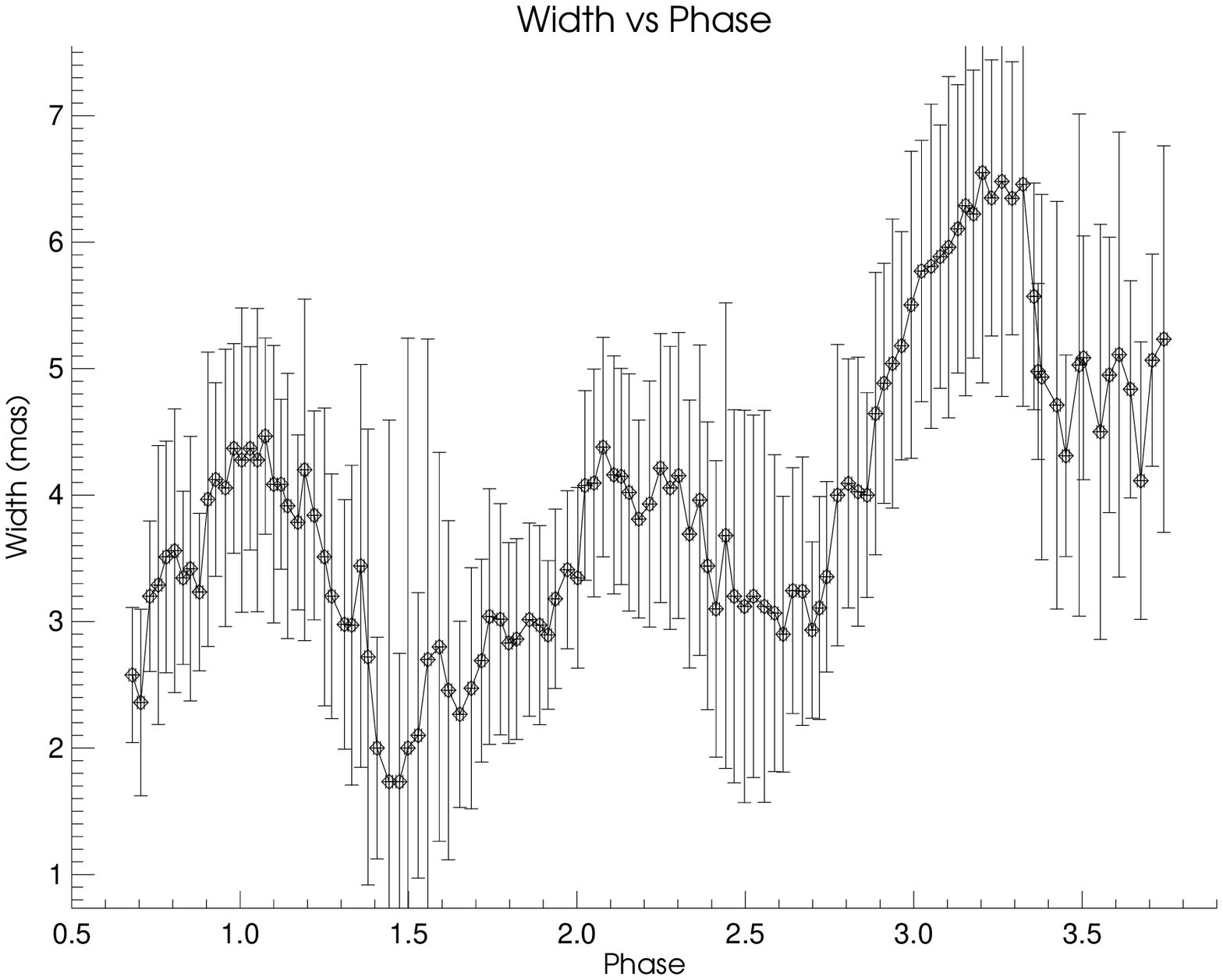}}\\
		\subfloat[]{\label{fig:wid_vs_rad}\includegraphics[trim=8mm 0mm 0mm 0mm, clip, scale=0.45]{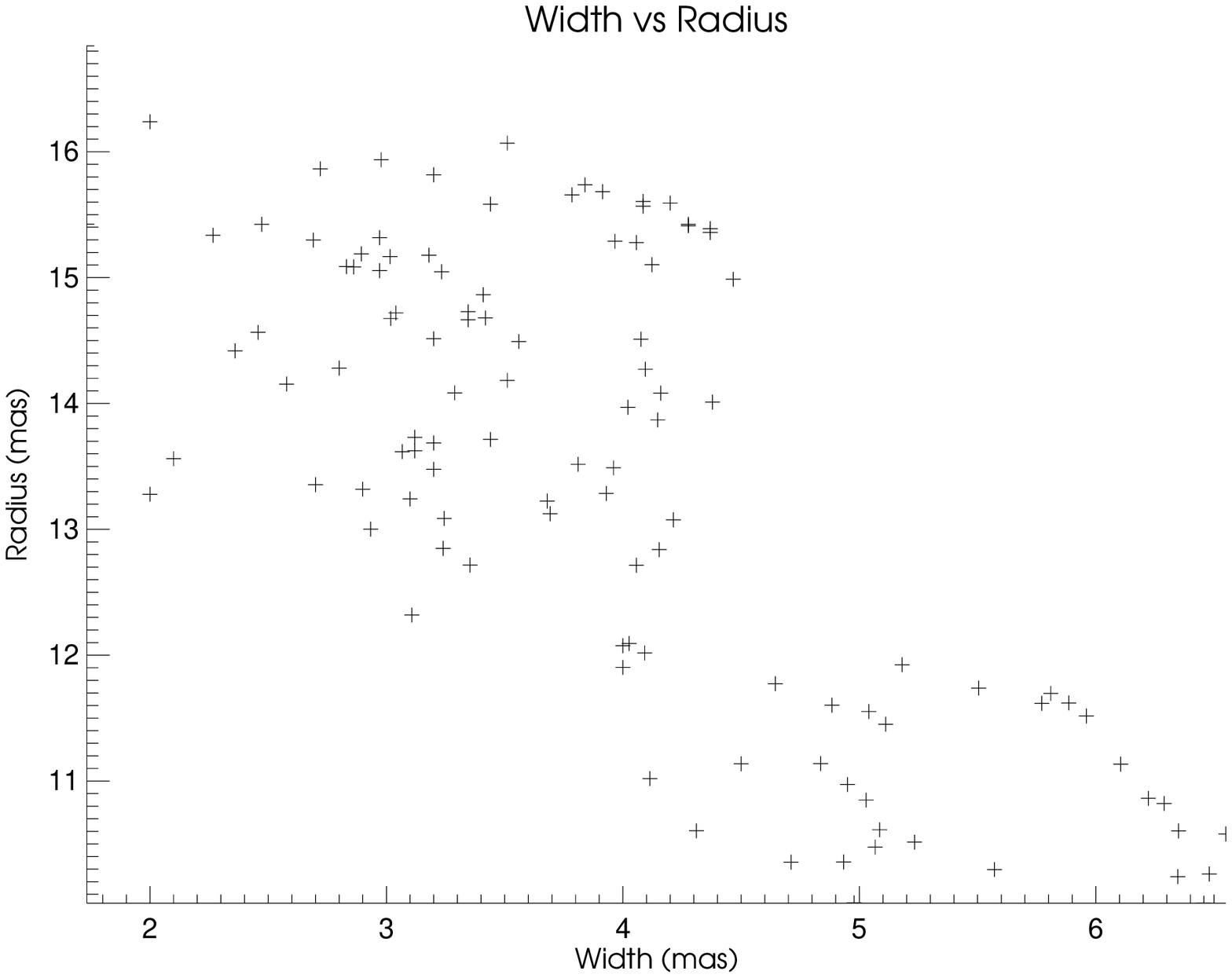}}\\
		\subfloat[]{\label{fig:wid_vs_int}\includegraphics[trim=8mm 0mm 0mm 0mm, clip, scale=0.45]{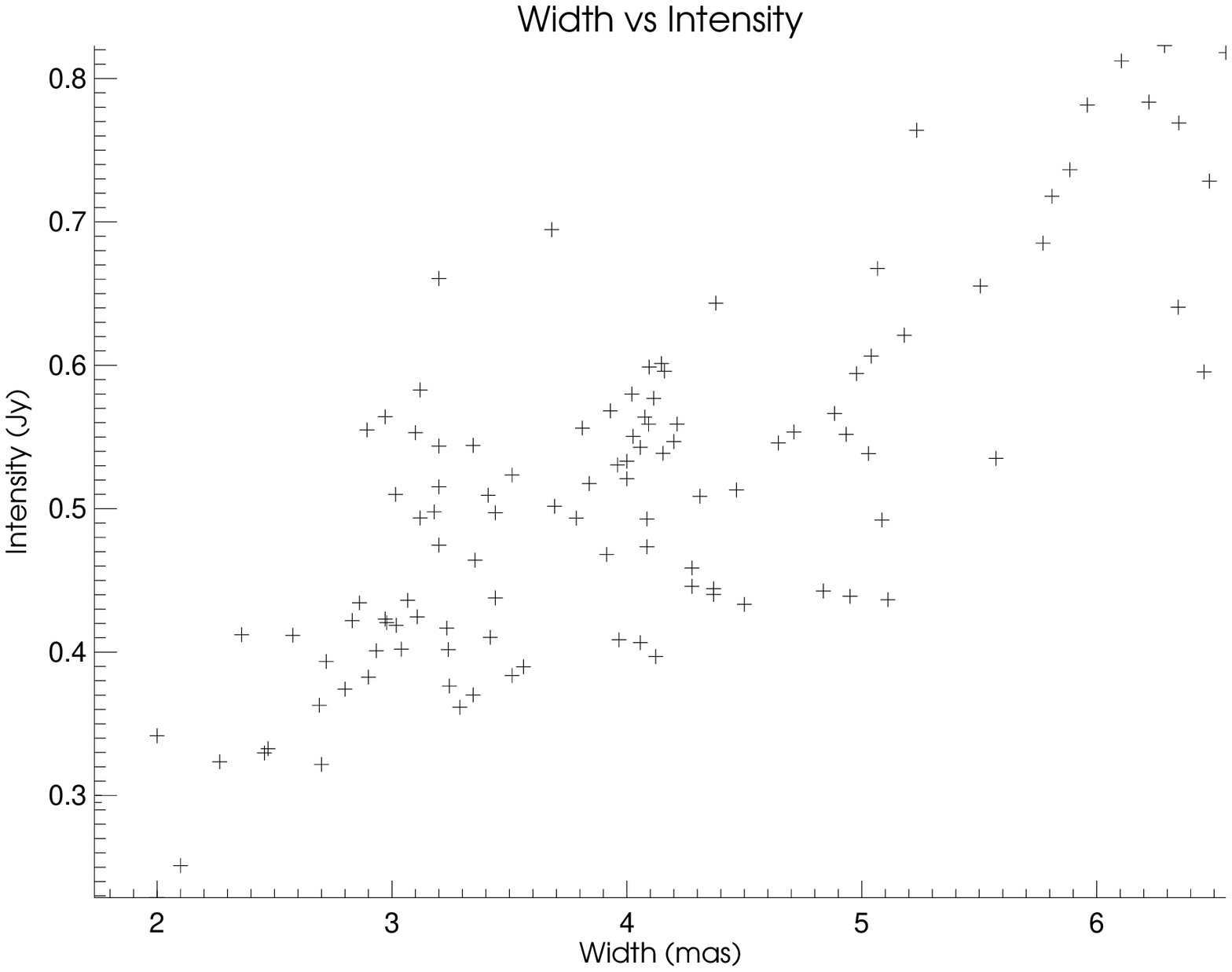}}\\
		\caption{(a) The average width of the ring plotted as a function of stellar 
		pulsation. (b) The width plotted against inner shell radius. (c) The width 
		plotted against sector-averaged intensity.}
		\label{fig:fig_4}
		\end{figure}

		\subsubsection{Morphological Evolution} 
		\label{sec:plot}
		The movie is 3.06 periods long but it is spread over 4 stellar cycles, 
		from $\phi$=0.68 to $\phi$=3.74. For convenience in the delineation of the 
		movie, we divide it into four parts, each covering an individual stellar cycle. 
		Features of interest (maser spots, either individual or in groups, with behaviour 
		that deviates from the overall kinematic behaviour)  are given a name to make it 
		easier to follow them, especially if they participate in more than one cycle.\\
		{\bf Cycle 1 \it($\phi$=0.64-1.00)}: The emission traces the perimeter of a 
		circle but with an obvious gap in the SE and a very weak NW quadrant that by 
		$\phi$=1.00 becomes almost totally void of emission. The E and S sectors start 
		as, and remain, the brightest sections of the ring. The emission is mainly 
		confined in spots and filaments; however, diffuse emission (Feature 1) is present
		since the first epoch and gradually dominates the very outer parts of the 
		aforementioned SE gap. The global motion is predominantly outflow throughout 
		this cycle; however, there are deviations from this pattern. An isolated 
		filament (Feature 2), not associated with the ring and at a distance from its 
		northern boundary, appears stationary and unaffected by the overall expansion. 
		Moreover, Feature 1 in the SE appears to be infalling and its inner parts 
		condense, forming some individual spots by $\phi$$\sim$0.98. At $\phi$$\sim$0.87 
		the Eastern part of the ring becomes very complex kinematically, with a group 
		of components (Feature 3) moving against the outflow and eventually plunging 
		toward the star. The SW quadrant appears to be moving faster than the NE 
		quadrant and at $\phi$$\sim$1.00 the overall structure resembles an ellipse 
		with its axis in the NE-SW direction and a protruding feature in the SE. \\
		{\bf Cycle 2 \it($\phi$=1.00-2.00)}: The western part of the ring becomes 
		brighter in this cycle but the eastern part remains the dominant source of 
		emission. At $\phi$$\sim$1.33 the SE-SW-NW hemicycle disappears and the 
		emission appears as a semicircle. Spots appear again in this region at 
		$\phi$$\sim$1.50 but the NW quadrant remains very faint. The main global 
		motion appears to be outflow again, albeit less pronounced than previously. 
		Feature 1 stops infalling and slows down until $\phi$$\sim$1.24 when it starts 
		moving outward, following the overall motion of the ring until it disappears 
		at $\phi$$\sim$1.71.  Feature~1 follows a peculiar path, culminating at 
		$\phi$$\sim$1.19 when it forms a secondary, much smaller ring which is 
		attached to the main structure. Feature 2 remains faint and nearly stationary 
		but, in an abrupt change of kinematic behaviour at $\phi$$\sim$1.17, it plummets 
		to the star and disappears at $\phi$$\sim$1.52.  Finally, Feature 3 continues 
		its infall until, at $\phi$$\sim$1.49, each spot appears to change direction 
		abruptly, making this feature the first {\it ricochet} observed in the movie. 
		The ring at this point resembles a figure-eight. The spots on Feature 3 follow 
		the ordered outflow once again until $\phi$$\sim$1.97, when they vanish and 
		the structure again resembles an ellipse, inclined on a NE-SW axis. \\
		{\bf Cycle 3 \it($\phi$=2.00-3.00)}: During this cycle, the projected ring 
		structure starts getting brighter reaching a maximum at $\phi$$\sim$2.15 at 
		which point it exhibits the most complete ring observed so far. Then it fades 
		again, reaching a minimum at $\phi$$\sim$2.71 with major regions of the S and 
		E sectors totally devoid of emission. Regarding the overall kinematics of the 
		ring, for the first time it appears that the predominant global motion is 
		contraction. The ring clearly reduces its radius, creating the smallest 
		apparent radial extent observed so far, at $\phi$$\sim$2.99. The shape of the 
		structure changes rapidly from an ellipse to a circle mainly due to a {\it 
		split} (Feature 4) starting at $\phi$$\sim$1.97 in the NE part. This specific 
		arc, almost as long as the NE quadrant, is clearly divided into two slivers. 
		While the outer one moves further outward disappearing at  $\phi$$\sim$2.46, 
		the inner component, still being a part of the whole structure, contracts 
		until $\phi$$\sim$2.18 when a new split occurs exhibiting the same behaviour.  
		The second apparent ricochet (Feature 5) appears at $\phi$$\sim$2.07, in the 
		northern part of the ring. Feature 5 follows the infall and becomes brighter 
		but at $\phi$$\sim$2.18 it reverses direction and stops, resisting the overall 
		inflow; it disappears at $\phi$$\sim$2.77. The overall circular symmetry of 
		the projected ring is broken at $\phi$$\sim$2.46 due to the development of a 
		formation of maser spots in the SW quadrant that  create a new disruption of 
		the ring structure. Circular symmetry is restored at  $\phi$$\sim$2.80.\\
		{\bf Cycle 4 \it($\phi$=3.00-3.74)}:  The ring during this cycle displays 
		different characteristics to the earlier cycles. It has the longest filaments 
		observed, bright and compact spots, and the ring is wider but covers a smaller 
		percentage of the perimeter compared to previous cycles. In this cycle the ring 
		continues contracting until $\phi$$\sim$3.32 when it reaches the smallest radial 
		extend observed throughout the whole movie. Then, the ring seems to undergo 
		expansion until $\phi$$\sim$3.58, at which time the emission appears to be at 
		its minimum with only the NE quadrant exhibiting substantial emission. The only 
		feature that defies the relatively uniform motion of the ring is a filament 
		(Feature 6) in the  N-NE part, that initially follows an infall motion and 
		at $\phi$$\sim$3.02 appears to bounce, continuing its motion to the opposite 
		direction. It then becomes brighter and elongated, breaks into smaller spots 
		at $\phi$$\sim$3.45 and finally disappears at  $\phi$$\sim$3.65.\\

	\subsection{Inner Shell and Width}	
			
	The overall behaviour and kinematics of the extended atmosphere of TX Cam can be 
	probed through the analysis of the global kinematics of the emitting region. 
	A major problem arises however from the fact that the structure of this region 
	changes with time and, although sometimes it conforms with simple geometrical 
	shapes, it is frequently confined in more complex or random formations. We took 
	advantage of a specific property of the emission in order to surmount this 
	difficulty; emission appears diffuse and scattered in the outer parts but it is 
	well defined in the inside. Thus, by defining the {\it inner shell boundary} 
	of the ring, we can then estimate an average value of the {\it inner radius} 
	and follow its change with time. For the calculation of the inner shell radius 
	and the width of the ring, we based our analysis in the technique developed 
	by \cite{diamond_1}. According to this approach, the ring is divided into 
	twelve sectors and the radial intensity function is calculated for each of 
	them. The inner shell radius for each sector is then calculated as the position 
	of the local maximum in the gradient of this function. Hence, the mean inner 
	shell radius is the average of all the sectors. 		
		
	In order to make the technique more sensitive to small-scale variations of the 
	inner shell radius, we divided the ring into 24 equal sectors (instead of 
	twelve) and calculated their radial intensity profile, the intensity as a 
	function of increasing distance form the centre of the image which is assumed 
	to be the centre of the star, in the same manner as in \cite{diamond_1}. 
	Although in most cases this profile is adequate to provide an accurate 
	estimation of the inner shell, there are features that can impose limitations 
	on this technique so, we developped a more global approach that does not 
	depend on the characteristics of individual frames. Different frames have 
	different noise levels thus, blanking the noise can cause an over or 
	underestimation of the inner shell boundary from frame to frame. On the 
	other hand, a constant cutoff level on all frames is also ill-advised, since 
	in noisier images artefacts would appear within the inner ring region from 
	the remaining unblanked noise and affect the determination of the inner shell 
	radius. Moreover, the inner boundary is not uniform but interspersed with 
	gaps due to the fragmented nature of the ring and there are occasions were 
	there is no gradient in the radial intensity that can be used to estimate 
	reliably the inner shell radius. Confusion can also be caused by splits or 
	the formation of a new ring; in these cases it can not be conclusively 
	determined whether emission inside the ring is short-lived and due to disappear 
	in the next frames or marks the birth of a new ring. To overcome these 
	obstacles, the radial intensity profiles from all 112 epochs, for each sector, 
	were concatenated in increasing order to create a three-dimensional plot, 
	showing the evolution of the intensity at different distances from the centre 
	over time. As seen in  Fig.~\ref{fig:surface},  here taken to show Sector 13, 
	the average intensity forms clearly distinguishable features in the plot. 
		
	We defined the inner boundary as the points that have an average intensity 
	three times the intensity of the off-source regions. To calculate the latter 
	quantity we averaged the intensities at large radii (24-25 mas) over time. 
	From this starting point, the inner shell boundary can then be traced by the 
	contour that corresponds to the aforementioned brightness. Features that last 
	for less than five consecutive epochs are not accepted as part of the main 
	ring and are excluded from the determination of the inner shell boundary. Each 
	point in the contour is characterised by its coordinates; its distance from 
	the centre of the star, the epoch of observations and its flux density, which 
	is the same for all the points of the contour (3$\sigma$). Since each contour 
	may provide two or more values of the inner shell radius for each epoch, the 
	point  with the lower distance from the centre of the star is accepted as the 
	inner boundary (the red crosses superimposed on the blue contour in 
	Fig.\ref{fig:surface}). The mean shell radius for each epoch is then calculated 
	as the average of all sectors. Fig.\ref{fig:ring} shows the results of this 
	technique in frame 55. Each sector is plotted over the total intensity image 
	as a triangle with its apex coincident with the star's centre and height equal 
	to the inner shell radius.	
		
	The {\it width} of the ring is also calculated with this technique. As 
	explained before, for each epoch, the point of the contour with the lower 
	value of distance from the centre of the star gives the inner boundary radius 
	of the sector. In the same manner, the upper value gives the outer boundary 
	radius and the difference between this lower and upper values give the width 
	of the ring at this sector. The average width is calculated as the mean of 
	all the sectors.

    \subsection{Variability}
    
		\subsubsection{Inner Radius}
		
		Fig. \ref{fig:radius} shows the average inner radius plotted with phase. With an 
		exception to the rupture at $\phi$$\sim$1.47, the continuous nature of the plot 
		suggests that the overall changes in the motion of the ring are smooth, in 
		accordance to the ordered flow observed in the movie. As described in Section 
		\ref{sec:plot}, from $\phi$$\sim$1.33 to 1.47, the SE-SW-NW part of the ring is 
		devoid of emission; the appearance of new spots in this region close to the star 
		at $\phi$$\sim$1.49 causes an abrupt change in the inner shell radius of the 
		masering shell, demonstrated by the rupture in the inner shell radius plot. The 
		ring appears to be expanding during the first cycle, with no prominent 
		contraction. This behaviour is somewhat atypical, as the plot reveals for the 
		other cycles,  the normal mode of motion is that of expansion during the first 
		half of the cycle and contraction during the second half. Across all cycles there 
		also appears to be a lag between the time at which the inner radius is at its 
		maximum or minimum with respect to the stellar phase maxima and minima. The extent 
		and duration of expansion or contraction also changes from cycle to cycle: as seen 
		in Table~\ref{table_2} the most intense contraction occurs between cycles 3 and 4, 
		when the inner radius decreases by 26.96\%. Contrary to what is observed during 
		the first cycle where there is no observable contraction and the ring expands by 
		18.98\%, the other cycles reveal an almost constant rate of expansion between 
		maxima and minima within each cycle of 13-14\%. The maximum and minimum inner 
		radii determined over the course of the three observed pulsation cycles are 16.84 
		and 10.03 mas respectively, that corresponds to 6.58 and 3.92 au from the centre 
		of TX Cam, calculated assuming a distance of 390 pc \citep{olivier_1}. 
		
		Fig. \ref{fig:distance} shows the average distance of the inner shell boundary 
		(the onset of the masering zone) from the surface of the star for a range of 
		different adopted sinusoidal pulsation amplitudes in stellar radius (0-30 \%). 
		For the radius of TX Cam we used the value from \cite{cahn_1} but adjusted to a 
		distance of 390 pc \citep{olivier_1} which is adopted in our study. Thus, the 
		stellar radius is R$_{\star}$=2.38$\times$10$^{13}$ cm at \phase=0.939 (4.08 mas 
		or 1.59 a.u.). The masering zone does not form at a constant distance from the 
		surface; from cycle to cycle, the SiO rings form progressively closer to the 
		stellar surface. A pulsation of 20\% yields a distance between 5.35-12.34 mas 
		(4.73-2.09 au or 2.81-1.24 R$_{\star}$). Even for the same pulsation phase, there 
		are big differences in this distance for different cycles. To demonstrate this, we 
		compare the distances at \phase\similar 0.68, 1.68, 2.68 and 3.68; the distance of 
		the inner shell from the surface of TX Cam is 9.56 mas (3.73 au or 2.21 
		R$_{\star}$), 10.84 mas (4.22 au or 2.51 R$_{\star}$ 8.43 mas (3.27 au or 1.95 
		R$_{\star}$) and 6.42 mas (2.50 au or 1.48 R$_{\star}$) respectively.

		Fig. \ref{fig:rad_vs_int} is a plot of the inner radius and the sector-averaged 
		intensity as a function of stellar pulsation. It appears that the two quantities 
		are correlated with rings located closer to the star being more intense.
	
		\begin{table*}	
		\center
		\begin{tabular}{cccccccccc}
		\hline
		\hline
        & \multicolumn{4}{c}{Radius} & &\multicolumn{4}{c}{Width}  \\
        \cline{2-5} 
        \cline{7-10}
		Cycle            & Minimum & Maximum   & \% Change & \% Change& &Maximum & Minimum   & \% Change & \% Change \\
		 {\it i}    & (mas) & (mas) & Min($i$)$\rightarrow$Max($i$) &Max($i$)$\rightarrow$Min($i$+1)&&(mas) & (mas) & Max($i$)$\rightarrow$Min($i$) &
		 Min($i$)$\rightarrow$Max($i$+1)\\
		\cline{1-5}
		\cline{7-10}
		1& 14.1538& 16.8411& 18.98\footnotemark[1] & -21.15\footnotemark[2]& & -\footnotemark[1] & 2.36 & - & 88.98\\
		2& 13.2786& 15.1782& 14.31& -20.03& & 4.46 & 1.73 & -61.21 & 152.61\\
		3& 12.1386& 13.7312&  13.12& -26.96& & 4.37 &  2.90 & -33.64 & 125.86\\
		4& 10.0290& 11.4514& 14.18& -8.18\footnotemark[3]& & 6.55 & 4.11 & -37.25 & - \\
		\hline
		\hline
		\end{tabular}
		\medskip
		\vspace{-0.2cm}
		\flushleft{1. Our observations start at $\phi$=0.64, thus the actual minimum of 
		the inner radius and the maximum width for the first cycle are unknown.}
		\vspace{-0.2cm}
		\flushleft{2. The transition from the maximum of the first cycle to the minimum 
		of the second is not smooth. The change given here occurs in an interval of 
		$\Delta \phi$=0.09, contrary to the other cycles were they occur smoothly and 
		last approximately half the stellar pulsation cycle.}
		\vspace{-0.2cm}
		\flushleft{3. Our data stop at $\phi$=3.74 so this value does not correspond to 
		the full change within the cycle.}
		\caption{The table shows the minimum and maximum values of the inner radius and 
		shell width and its percentage change from consecutive minima and maxima.}
		\label{table_2}
		\end{table*}	

		\subsubsection{Width}
		Fig. \ref{fig:width} plots the width of the projected maser shell as a function 
		of stellar pulsation phase. This plot clearly shows that the width of the ring 
		appears to follow the pulsation of the star. However, the minimum width appears 
		at a lag relative to the minimum in the optical, as so does the maximum. The 
		values of the width range from 1.73 to 6.55 mas (0.68-2.55 au or 
		0.40-1.51 R$_{\star}$). The maxima and minima widths can change from period to 
		period. During the maxima of the second and third cycle (our data do not cover 
		the maximum of the first maximum observed) the ring appears equally wide 
		($\sim$4.37 mas, 1.70 au or 1.01 R$_{\star}$), on the fourth it is increased to 
		6.55 mas (2.55 au or 1.51 R$_{\star}$). The width minima demonstrate similar 
		properties with not very wide rings for the first and second minima (2.36 and 
		1.73 mas, 0.90 and 0.68 au or 0.54 and 0.40 R$_{\star}$ respectively), but 
		increasing to 2.90 and 4.11 mas (1.13 and 1.61 au or 0.67 and 0.95 R$_{\star}$) 
		for the third and fourth cycles respectively. Fig. \ref{fig:wid_vs_rad} shows that 
		wider rings tend to form closer to the star. Fig. \ref{fig:wid_vs_int} plots 
		intensity against shell width. It appears that the stronger the emission, the 
		wider the masering zone. This dependence explains the different values of the 
		width at maxima and minima, since stronger cycles should have wider masering 
		zones. 

        \subsubsection{Flux density and Lags}
		Fig. \ref{fig:fig_4} shows the light curve of TX Cam (in red) for the period 
		corresponding to the length of the movie, provided by AAVSO. The flux density 
		of the SiO masers (in blue) within the ring is also plotted as a visual aid to 
		reveal any correlation between the two quantities and compare their properties. 
		It is obvious that both visual magnitude and radio flux density  follow the 
		stellar pulsation but there are some very distinctive differences. 
	
		The visual magnitude and radio flux density are uncorrelated in their detailed 
		properties. The maxima and minima for the optical appear almost constant in all 
		cycles, thus their ratio i.e. their contrast, remains constant. On the other hand, 
		the radio flux densities at minima and maxima progressively increase, having a 
		variable contrast . Moreover, there appears to be a delay between the radio and 
		the optical. In order to examine this in more detail, we fitted polynomials to the 
		optical curve, the 43 GHz flux density and the width of the ring. From the fitted 
		polynomials we calculated the lag of the radio flux density and the width with 
		respect to the optical magnitude. The results are summarised on Table \ref{table_3}. 
		On average, the radio peaks appear delayed (around 13\%) with respect to the 
		optical, as also found by other studies \citep{alcolea_1,pardo_1}; the same 
		applies to the maximum widths. 
	
		\begin{table*}	
		\center
		\begin{tabular}{ccccccccccccc}
		\hline
		\hline
        & \multicolumn{2}{c}{Optical} & &\multicolumn{4}{c}{Radio} & &\multicolumn{4}{c}{Width}  \\
		Cycle            & Maximum & Minimum & & Maximum & Lag & Minimum & Lag & &Maximum & Lag & Minimum & Lag \\
		\cline{2-3}
		\cline{5-8}
		\cline{10-13}
		1& 0.975& 1.462& & 1.094& 11.9 & 1.550 &  8.8  &  & 1.020 & 4.5   & 1.527 & 6.5\\
		2& 2.016& 2.526& & 2.159& 14.3 & 2.636 &  11.0&  & 2.116 & 10.0 & 2.552 & 2.6\\
		3& 3.021& 3.525& & 3.176& 15.5 & 3.569 &  4.4  &  & 3.220 & 19.9 & 3.616 & 9.1\\
		\hline
		Avg. &     &           & &           & 13.90 &             &  8.06 & &            &  11.46 &         & 6.06\\
		\hline
		\hline
		\end{tabular}
		\caption{The table shows the phases at which the minima and maxima occur according 
		to fitted polynomials to the visual magnitude, the radio flux density and the 
		width of the ring. The lag is given with respect to the optical as a percentage 
		of the stellar pulsation period.}
		\label{table_3}
		\end{table*}	

		\begin{figure}
		\centering
		\subfloat{\label{fig:radius2}\includegraphics[trim=20mm 0mm 0mm 0mm, clip, scale=0.55]{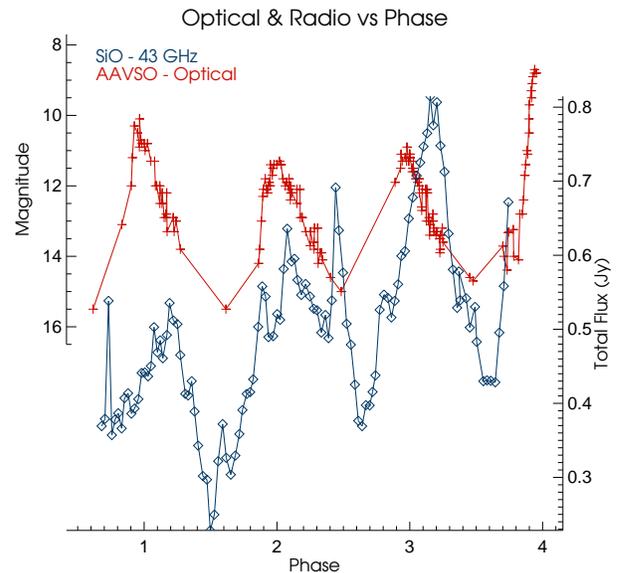}}
		\caption{The visual magnitude (provided by AAVSO in red) and the radio flux 
		density at 43 GHz in blue, plotted as a function of the stellar pulsation phase.}
		\label{fig:fig_4}
		\end{figure}

\section{Shock Waves}
\label{sec:shock_waves}

\begin{figure*}
\centering
\subfloat[]{\label{collage_1}\includegraphics[trim=0mm 0mm 0mm 0mm, clip, scale=0.45]{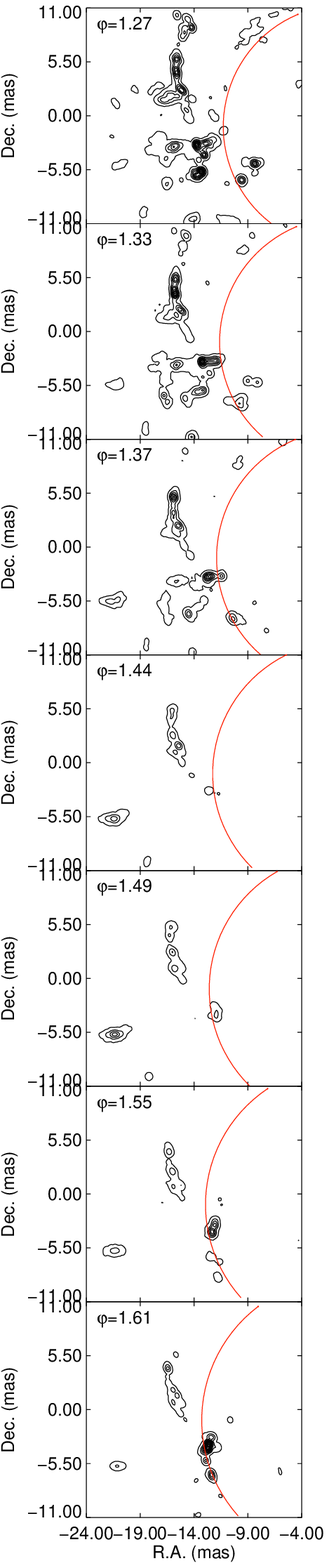}}
\subfloat[]{\label{collage_2}\includegraphics[trim=0mm 0mm 0mm 0mm, clip, scale=0.45]{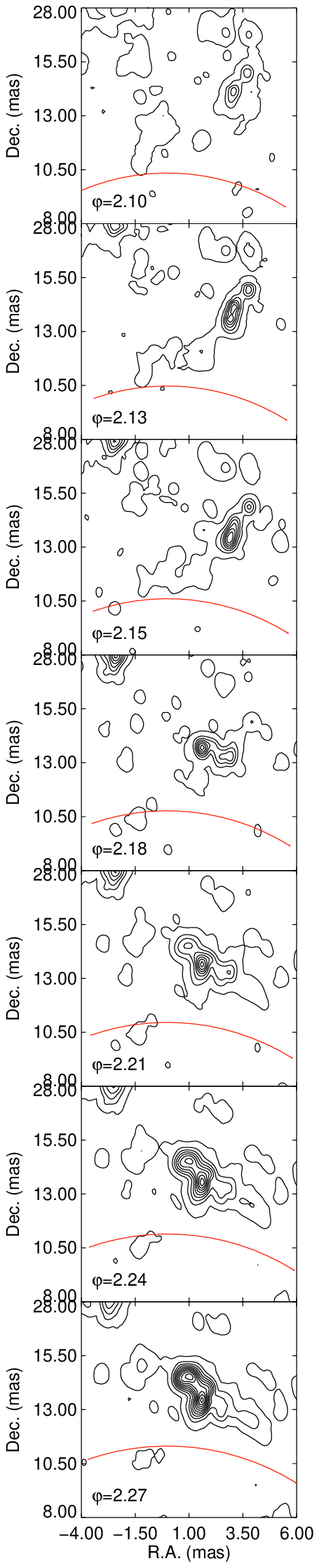}}
\subfloat[]{\label{collage_3}\includegraphics[trim=0mm 0mm 0mm 0mm, clip, scale=0.45]{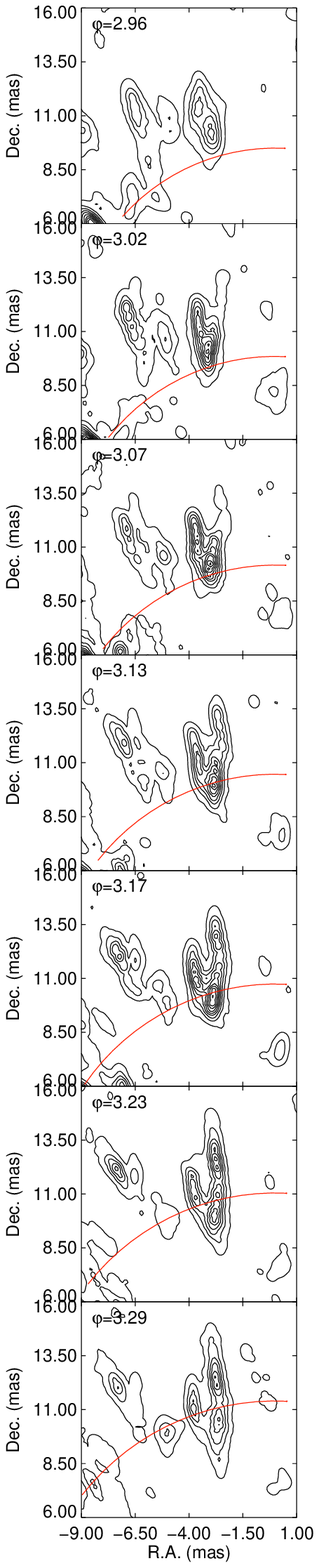}}
\subfloat[]{\label{collage_4}\includegraphics[trim=0mm 0mm 0mm 0mm, clip, scale=0.45]{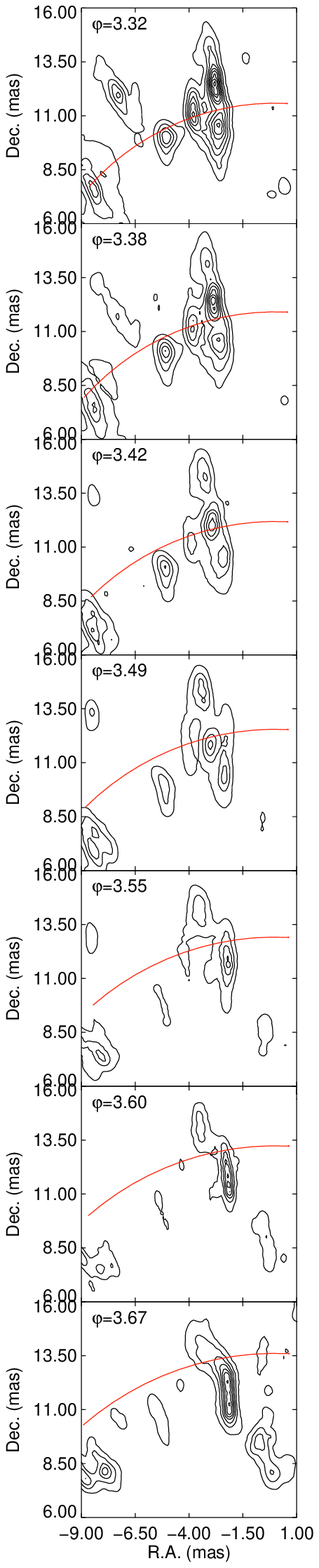}}
\subfloat[]{\label{collage_5}\includegraphics[trim=0mm 0mm 0mm 0mm, clip, scale=0.45]{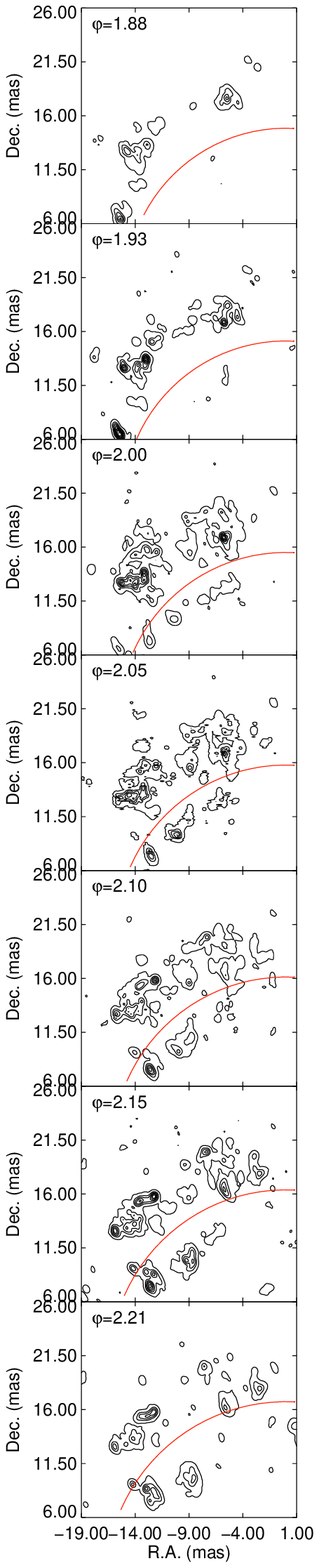}}
\subfloat[]{\label{collage_6}\includegraphics[trim=0mm 0mm 0mm 0mm, clip, scale=0.45]{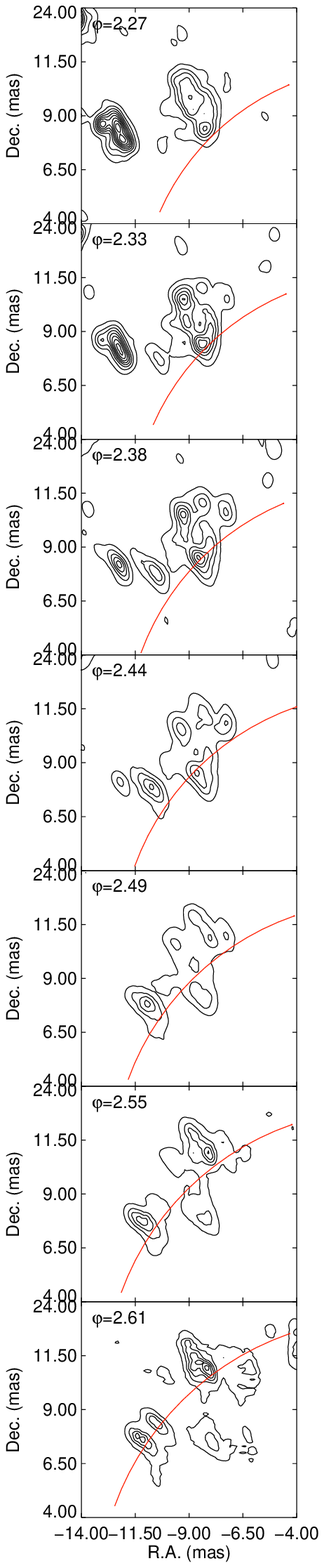}}
\subfloat[]{\label{collage_7}\includegraphics[trim=0mm 0mm 0mm 0mm, clip, scale=0.45]{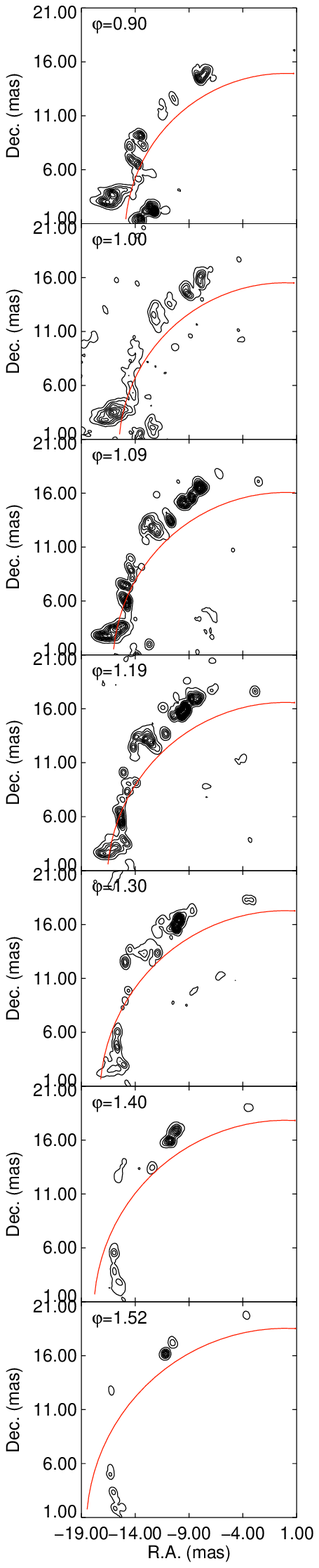}}
\subfloat[]{\label{collage_8}\includegraphics[trim=0mm 0mm 0mm 0mm, clip, scale=0.45]{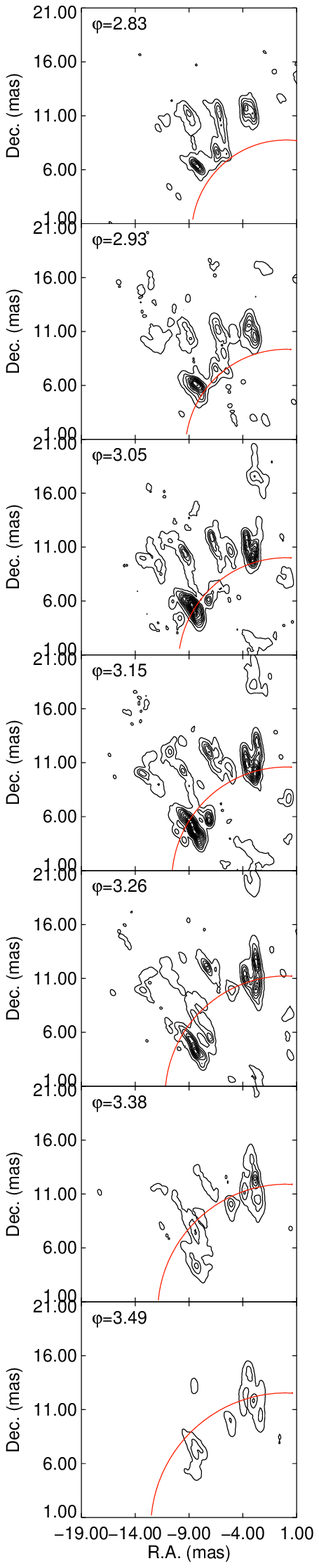}}
\caption{Each column is a collage of consecutive epochs, zoomed in to a feature of 
interest. In all the frames, the proposed shock is plotted as a red circle. The centre of 
the star is the centre of the circle. {\bf (a) Feature 3}. A group of maser spots starts 
falling inwards; it appears that the encounter of these spots with the shock front 
results into a change of the direction of their motion.  {\bf (b) Feature 5}. The motion 
of this feature appears to be opposite to the global kinematic behaviour. However, as it 
meets the shock front, it changes direction following the overall outflow. {\bf (c) and 
(d) Feature 6}. According to the proposed shock wave, this feature appears to be 
infalling. As it encounters the shock front, the feature changes its direction of motion. 
Then, as the shock propagates through the feature, it elongates it and eventually, breaks 
it down to into smaller spots. {\bf (e) and (f) Feature 4}. This is the first split 
observed. The addition of the shock front reveals that the prominent double arc is not a 
split but a condensation of post-shock material that starts falling toward the star. The 
shock appears to be pushing the pre-shock material as it moves out. However, in column 
(f), which is an additional zoom into this infalling material, it appears that the new 
shock front splits it into two new but smaller arcs. {\bf (g) and (h)} A comparison of 
the effects of the proposed shock. During the first cycle the shock appears to be pushing 
the material outwards. In contrast, in the last cycle observed, the shock seems to be 
penetrating the masering zone.}
\label{fig_collage}
\end{figure*}

It is generally believed that in Miras a shock wave leaves the stellar surface once per 
stellar cycle, when the star is at its maximum \citep{bowen_1,humphreys_1}. The effects 
of the passage of a shock through the extended atmosphere, where the SiO masers are 
located, could be visible in a number of ways. For example, a shock could split the ring 
to its pre- and post-shock components or make infalling material bounce back, forcing it 
to follow the shock's motion. In order to study the effects of a shock wave in the 
masering zone and identify which component  kinematics can be attributed to its passage, 
we created a second version of the movie, this time overlaying a shock wave.

\begin{figure*}
	\subfloat[]{\label{fig_movie_imax}\includegraphics[trim=9mm 0mm 0mm 30mm, clip, scale=0.45]{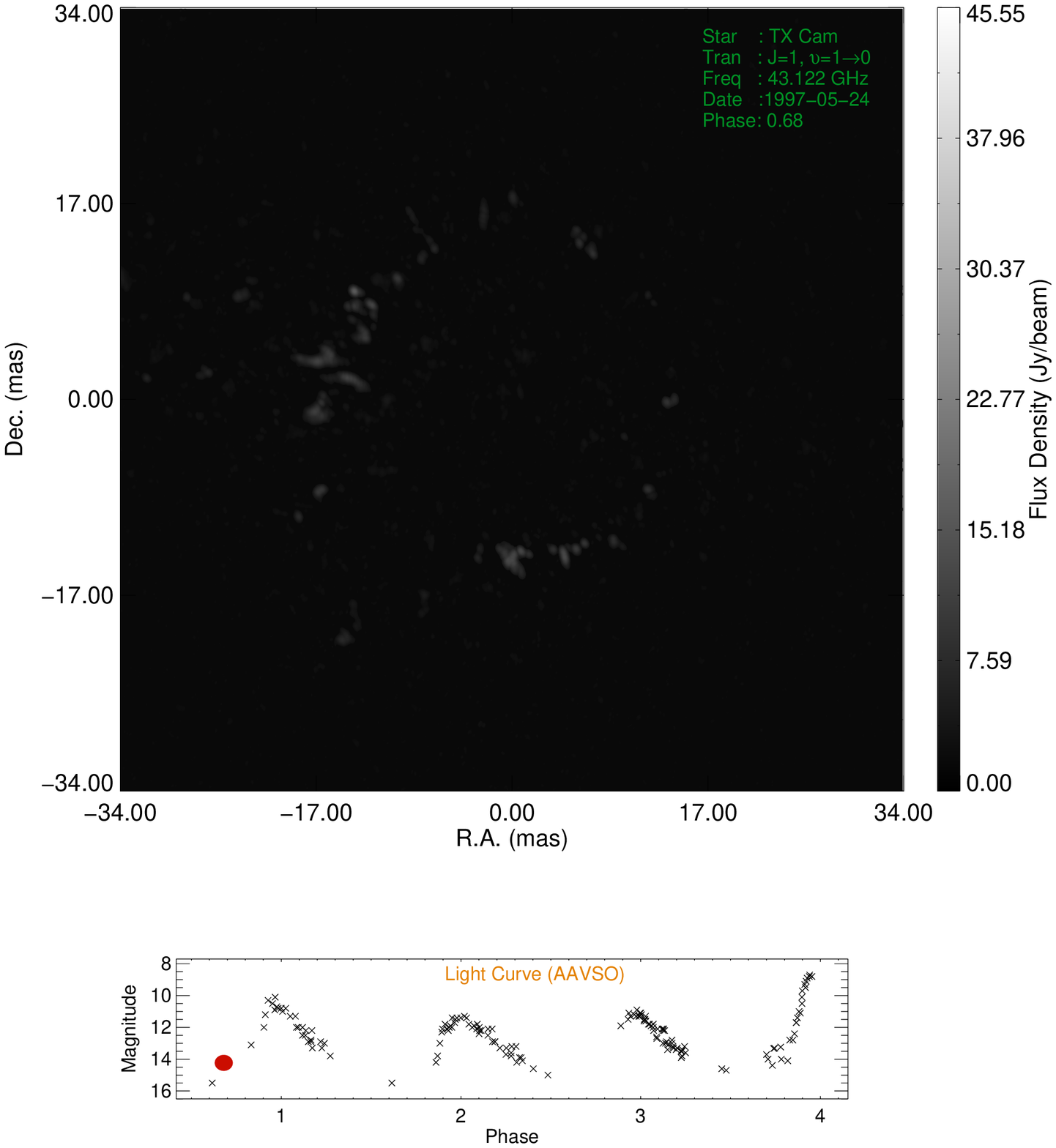}}
	\subfloat[]{\label{fig_movie_star}\includegraphics[trim=0mm 0mm 0mm 30mm, clip, scale=0.45]{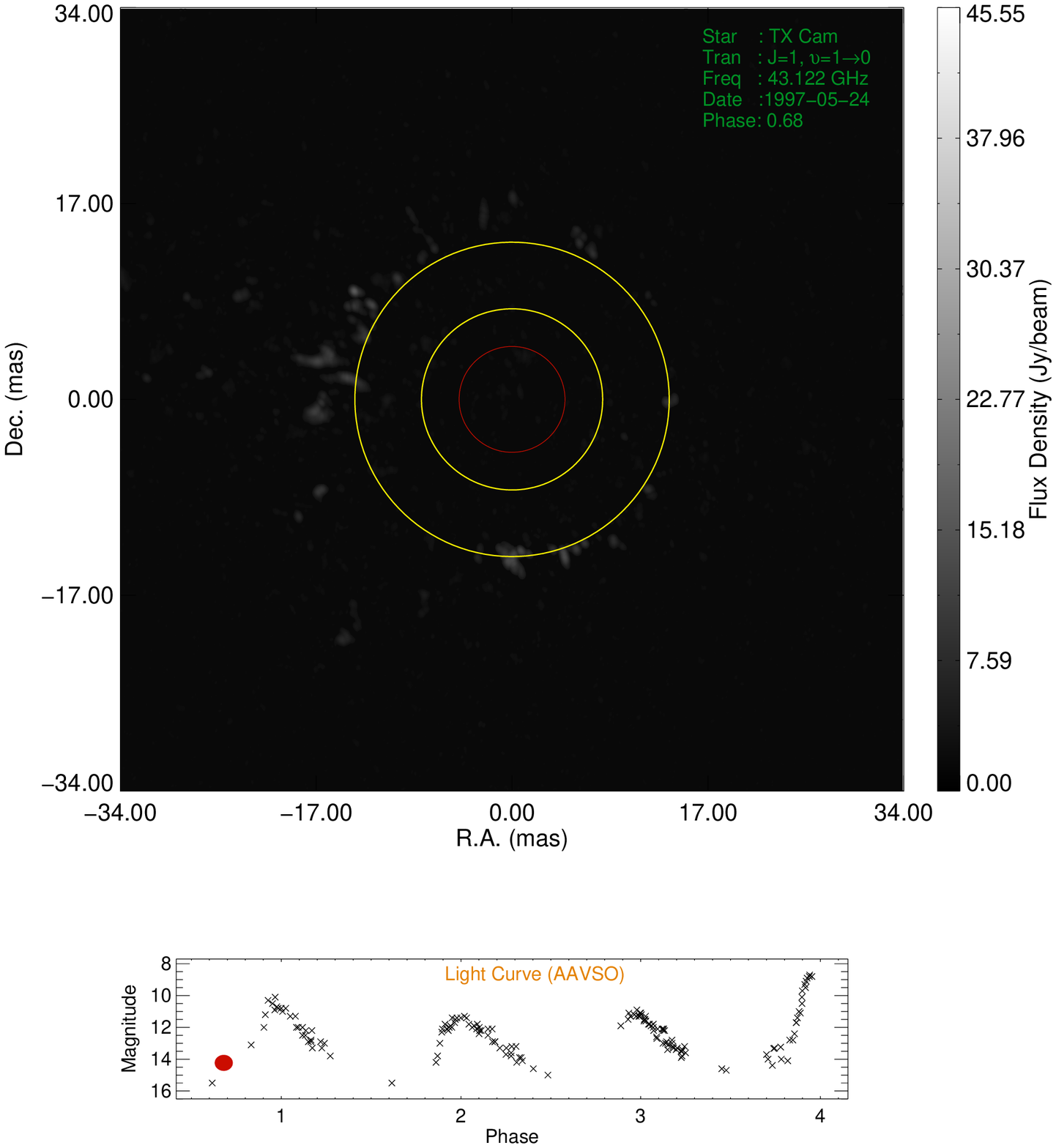}}
	\caption{(a) The first frame of the 112-frame movie, showing the evolution and global 
	kinematics of the SiO masers toward TX Cam. The lower panel is the optical light 
	curve, provided by AAVSO. The red dot follows the phase. (b) Same as in 
	Fig.~\ref{fig_movie_imax}, but with the addition of the proposed shock (yellow 
	circles) and the pulsating star (red circle). Both movies are provided as online 
	supplementary material.}
	\end{figure*}

For the radius of TX Cam we used the value from \cite{cahn_1} and also assumed a 20\% 
change in the stellar radius due to pulsation (as motived by the data of \cite{pettit_1} 
and used in \cite{reid_2} and also observed by \cite{wittkowski_1}). Additionally we 
adopted a shock velocity of 7 \kms, in line with the value of the initial velocity 
calculated for the inner shell boundary by \cite{diamond_1} and \cite{gonidakis_1}. With 
these parameters we were able to calculate the radius of a uniform shock wave, leaving 
the stellar surface at maximum light. Once this information was also included in the 
frames, they were converted into a movie showing the propagation of the proposed shock. 
The movie is provided as online supplementary material and Fig.~\ref{fig_movie_star} 
shows its first frame.

Generally the shock appears to be intensifying the maser activity (Fig.~\ref{collage_8}) 
or even creating maser spots as it passes through the masering zone; there are however 
occasions that this is not the case. Starting at $\phi\sim$ 0.64, the position of the 
shock coincides with most parts of the ring. The NE region however appears to be more 
curved than the spherical shock so that, as it propagates, it appears that it is pushing 
the material away from the star rather than penetrating this region. As a result of the 
curvature of this region, the shock reaches the east and north sides first, gradually 
gaining ground and at $\phi\sim$1.71 it passes over the last spot of the NE tip 
(Fig.~\ref{collage_7}). Once overtaken, the spots fade away; however, on the eastern 
part, the material that the shock transits starts falling toward the star forming 
Feature 3. 

The movie can also reveal whether the peculiar kinematics of the features described in 
\S~\ref{sec:plot} are caused by the transition of a shock. Fig.~\ref{collage_1} shows 
Feature 3, the first ricochet observed. As can be seen, as the spots infall they meet the 
shock front and, at \phase\similar 1.44, they start changing the direction of their 
motion. The position of the shock at this phase coincides with the position where the 
change in the kinematics of the maser spots is observed. The addition of the shock front 
in the movie reveals also the true nature of the first split observed in Feature 4. It 
appears that the inner arc is not created by a split in the perimeter of the ring due to 
the shock but is actually infalling material, condensed 
in the post-shock region. The second split of this feature coincides perfectly with the 
passage of a shock through the ring, that splits it into two distinctive regions 
(Fig.~\ref{collage_5} and \ref{collage_6}).  
Finally, both features 5 and 6 show an agreement between their kinematics and the position 
of the shock front. More specifically, Feature 5 seems to be infalling and then, as it 
meets the shock, it appears to become stationary and brightens as the shock propagates 
through its volume and then dissipates (Fig.~\ref{collage_2}). The shock has a more 
dramatic influence in the structure of Feature 6; as it falls inward, it meets the
shock that then drags it outward, breaking it up into smaller spots that eventually 
disappear (Fig.~\ref{collage_3} and \ref{collage_4}).

\section{Discussion}
	\subsection{Variability}
		\subsubsection{Maser Distribution}
		The most prominent overall property of the emission is its mostly circular 
		distribution. The maser phenomenon is preferentially observed in directions along 
		the line of sight, where the column density of SiO and velocity coherence are 
		higher. In the extended atmospheres of AGB stars, where molecular species occupy 
		shells around the star with radial acceleration due to shocks, this occurs in the 
		tangent of the shell. This ``tangential amplification pattern" is responsible for 
		the ring-like structure of the masering areas. The cover factor, i.e the 
		completeness of the ring, changes with the stellar pulsation and it appears that 
		rings are formed preferentially near stellar maxima while, near minima, masers 
		occupy a smaller percentage of their shell perimeters. Moreover, rings at the 
		same phase do not exhibit the same characteristics. An inspection of the frames 
		at phases 0.75, 1.74, 2.74 and 3.74 reveals that, although all correspond to the 
		same pulsation phase of the ring, they can be significantly different. At 
		$\phi$$\sim$0.75 the SE and NW arcs are missing, at $\phi$$\sim$1.74 the SE arc 
		is faint but present, at $\phi$$\sim$2.74 a gap is located in the S sector, with 
		the E-SE-S-SW part of the ring very faint while at $\phi$$\sim$3.74 the SW arc is 
		missing with the SE-S-SW-W arc almost absent. These inconsistencies indicate that 
		conditions can vary greatly around the star and local phenomena might be of 
		particular importance. Moreover, the model of a spherical shell and a perfectly 
		formed shock wave are clearly too simplistic to explain the variations we observe. 

		The variability in the distribution of the masers is not only observed in the 
		degree of ring completeness but also in its radius relative to the star and its 
		width. The study of the inner shell radius and the width revealed that the 
		masering zone is neither located at a constant distance from the centre of the 
		star nor does it have a constant width. It is a very dynamic region, revealing 
		complex kinematics and its characteristics can change from cycle to cycle. The 
		radius of the inner shell usually increases from minima to maxima and decreases 
		from maxima to minima, causing the whole structure to alternately expand and 
		contract. An exception to this behaviour is the first cycle observed, where the 
		ring is continuously expanding. More importantly, the distance at which the ring 
		forms is not correlated with the stellar phase; at the same phase during 
		different cycles, rings can be located at different distances. The same appears 
		to apply to the width of the ring; although it follows the stellar pulsation with 
		a lag, it does not have the same characteristics at the same phases in different 
		cycles. Both the inner radius and the width appear to be correlated with the SiO 
		flux density and more intense rings are formed closer to the star and are wider.  
		This also implies that wider rings are located closer to the star.    

		\subsubsection{Flux Densities}
		The nature of the project allows us to study the flux variability of SiO masers 
		not only in the short-term within a cycle, but also in the long-term, from one 
		cycle to another. In line with \cite{gonidakis_1}, the sinusoidal pattern of the 
		flux densiy of the \emission emission reveals that it is correlated with the 
		stellar phase \citep{hjalmarson_1}. On the other hand, the visual magnitude and 
		radio flux density, appear to be uncorrelated in detail. Although all optical 
		maxima appear to be equally strong, the 43 GHz flux densities at maxima appear to 
		be increasing from cycle to cycle. The same applies to the minima, thus the 
		contrast, i.e. the ratio of fluxes in consecutive maxima and minima, is not 
		constant. 
	
		There is a clear, but not necessarily constant, lag between the visual magnitude 
		and the 43 GHz flux density in all the cycles observed \citep{cho_1}. A shock is 
		believed to be created near the stellar surface once per stellar maximum (see 
		$\S$\ref{subsec_shock_waves}) and, if SiO masers are collisionally pumped, then 
		the propagation of this shock through the masering zone can enhance maser 
		activity; the observed lag could be attributed to the time needed for the shock 
		to reach the masers. However, the strong correlation of the SiO flux densities 
		with the 8 $\mu$m IR radiation \citep{bujarrabal_1,jewell_1}, is a property that 
		can support the radiative pumping scenario. Thus, the lag itself cannot be 
		conclusive on the pumping mechanism of SiO masers.

	\subsection{Shock Waves}	
	\label{subsec_shock_waves}
	The idea of a shock wave in starsÕ atmospheres was firstly suggested by 
	\cite{merrill_1} in order to explain the big differences in the spectra of R Cyg.
	\cite{gorbatskii_1} produced the first theoretical model to show that emission 
	lines can be associated with shock waves that heat the atmosphere while they 
	propagate outwards through a low density region surrounding a long period star.
		
	There are a number of features in the movie that are dominated by peculiar 
	kinematics and we attempted to associate them with the existence of a shock wave 
	permeating the extended atmosphere of TX Cam. As discussed in 
	section~\ref{sec:shock_waves}, we overlayed in the movie a proposed shock wave, 
	leaving the stellar surface once per cycle at the maximum. As can be seen in
	Fig.~\ref{fig_collage} the changes in the kinematics and the morphology of these 
	features can be explained by the proposed shock wave, since they occur when the 
	shock front encounters the maser spots.
	
	The proposed shock appears to interact with the material in a number of ways. 
	Predominantly the shock appears to permeate the masering zone, increasing the 
	intensity of the maser spots (Fig.~\ref{collage_8}). This could be the result of 
	the increased rate of collisions due to the passage of the shock and evidence of 
	the collisional pumping mechanism responsible for SiO masers. However, during the 
	first cycle, the shock appears to be affecting only the kinematics of the ring,  
	being the driving force that pushes the masering material to the furthest position 
	observed throughout the movie (Fig.~\ref{collage_7}). From our data we cannot 
	conclude that the SiO masers are purely collisionally pumped; however, as in 
	\cite{cho_1}, it appears that collisions play an important role in the pumping, 
	especially around \phase\similar0.2.

	\subsection{Bipolarity?}
	\label{sec:bipolarity}
	
	\begin{figure}
	\centering
	\includegraphics[trim=0mm 0mm 0mm 0mm, clip, scale=0.43]{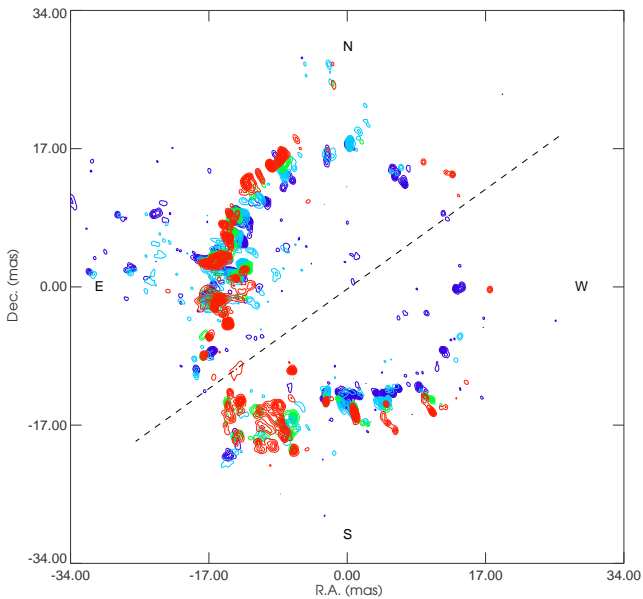}
	\caption{A figure showing contours maps of epochs BD46A, E, K and P (purple, blue, 
	green and red respectively). The maps are overplotted to show the differences in 
	expansion along the NE-SW and SE-NW axes. It is apparent that along the latter, the 
	flow of material is faster. }
	\label{fig:fig_5}
	\end{figure}
		
	The fit of a spherical shock wave revealed that the masering shell does not expand 
	uniformly. A characteristic example is the expansion during the first cycle, where the 
	masering ring appears to be more curved than the assumed perfectly circular shock. 
	Could that be a hint of bipolar outflow around TX Cam? In order to examine this 
	possibility we overlayed four different frames separated by five epochs to exaggerate 
	the differences, as seen in Fig.~\ref{fig:fig_5}. It is obvious that there are 
	differences in the outflow, with spots in the NE-SW axis being stationary, while the 
	SE-NW outflow appears to be the most intense. Another important observation is that 
	the maser emission itself is manifestly different in these two directions. In the 
	NE-SW direction there are clear gaps in the ring and emission is mainly diffuse. In 
	contrast, the SE-NW spots are much brighter and compact. All the above constitute 
	evidence of bipolar outflow around TX Cam. 
	
\begin{figure}			
\centering
\includegraphics[trim=8mm 0mm 0mm 0mm, clip, scale=0.45]{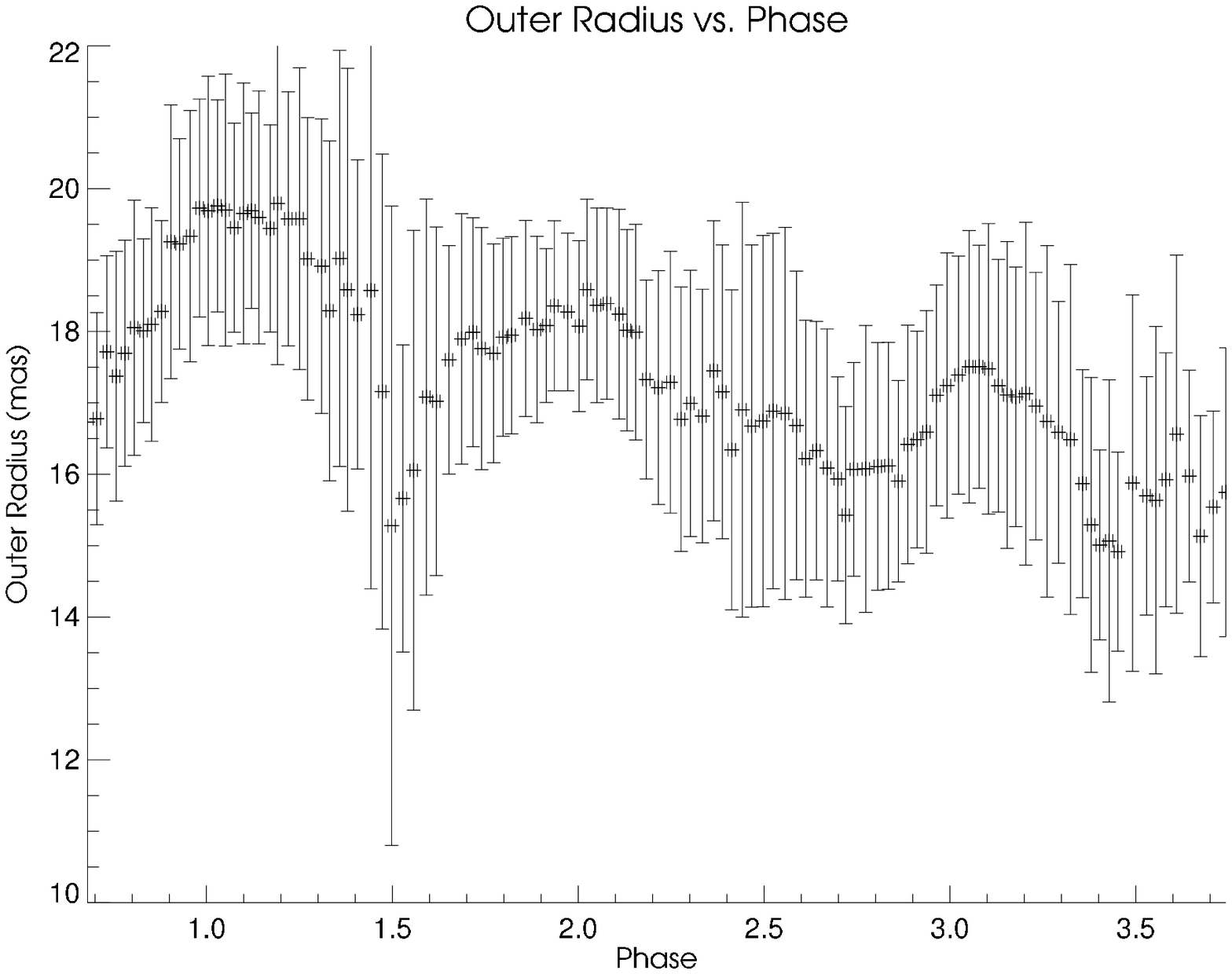}
\caption{This diagram shows the outer ring of the masering shell as calculated by the 
addition of the previously estimated quantities of the inner radius and the width.}
\label{fig:outer}
\end{figure}

	It appears that a shock with velocity of 7 \kms (as modelled by \cite{reid_2}) is in 
	good agreement with our data. According to our approach, the outer part of the ring 
	is dominated by diffuse emission and was difficult to define robustly. However, we 
	have calculated the radius of the inner boundary and the width of the ring so, their 
	sum provides an estimate of the outer ring radius. The shape of the plot for the 
	outer radius reveals a very important property of the masering shell: it follows 
	the pulsation of the star with a small lag and it does not appear to be dependent on 
	the overall kinematics of the inner shell, imposing an outer limit in the masering 
	zone that is dependent only on the phase (Fig.~\ref{fig:outer}). So, despite the lack 
	of contraction during the first cycle, the average outer boundary of the ring is 
	contracting; the constant expansion of the inner boundary is compensated by a thinner 
	shell. This difference in the behaviour between the inner and outer radii of the ring, 
	with the outer ring appearing unperturbed relative to its inner counterpart, might be 
	an indication that the shocks at this distance are damped. This is in agreement with 
	the result by \cite{reid_2}  stating that any periodic shocks or disturbances near 
	2$R_{\star}$ probably propagate outward with velocities less than 5 \kms and are 
	mostly damped.
	
	\subsection{Projection Effects}
 	For the values of stellar radius and distance adopted here, one can calculate the 
 	distance of the masering zone from the surface of TX Cam. The unknown in this case is 
 	how much does TX Cam's radius change from minimum to maximum. We calculated the 
 	distance of the ring from the surface of the star for several modes of pulsation, 
 	from 0-30\%, and the surface in Fig.~\ref{fig:distance} shows the result of this 
 	approach. Mid-infrared interferometry of S Ori with the VLTI/MIDI (Very Large 
 	Telescope Interferometer/MID-infrared Interferometric instrument) at four epochs, 
 	revealed that the photospheric radius shows a significant phase-dependent size with 
 	amplitude of $\sim$20\%, that is well correlated in phase with the visual lightcurve 
 	\citep{wittkowski_1}. As discussed in $\S$\ref{subsec_shock_waves}, for a 20\% change 
 	in the stellar radius, the distance of the inner shell from the surface of TX Cam 
 	ranges from 5.35 to 12.34 mas. Additionally, during the same phase at different 
 	cycles, this distance can vary greatly; the inner shell in the first cycle at 
 	\phase\similar0.68 is located 1.67 times further than at \phase\similar3.68.
 
 	The previous result is somewhat peculiar, since it suggests that a significant change 
 	in the conditions on the extended atmosphere occurs in such a way, that the masering 
 	zone is systematically positioned closer to the star form one cycle to the next. 
 	Our results also suggest that the ring becomes brighter and thicker as we progress in 
 	cycles. These results are independent on the distance adopted in our 
 	study. The answer seems to be hidden in the velocity structure of the ring and the 
 	individual maser spots (an explanation of the aforementioned peculiarities will be 
 	given in the next paper of this series, where the velocity information will be 
 	included in our analysis). As previously observed by \cite{yi_1} and \cite{gonidakis_1}, 
 	filaments exhibit a gradient along their axis, with values reaching the systemic 
 	velocity as we move along their axes toward the centre. Moreover, an examination of 
 	the radial velocities of the ring reveals that each cycle is very different 
 	kinematically along the line of sight. The ring during the first cycle is dominated by 
 	maser spots with velocities close to the systemic, but as we move in phase, spots 
 	appear systematically more blue- or red-shifted. This implies that projection effects 
 	are becoming more significant as we progress in phase. However, this also suggests 
 	that if projection is substantial, then a 3-dimensional study of the kinematics should 
 	be more appropriate in order to reveal the properties of the masering region.

	\subsection{Moving spots or change in conditions?}
	The maser spots appear to be moving, most of the time in an orderly manner, 
	contributing to the global kinematic behaviour of expansion or contraction. But are 
	we really witnessing the actual flow of material or is the emission tracing the 
	changes in the physical conditions? As has been previously observed, the kinematics 
	of the maser components follow ballistic motions, ruling out the latter scenario 
	\citep{diamond_1}. Additionally, there are components that exhibit kinematic 
	properties consistent with motions along magnetic field lines \citep{kemball_3}. 
	There are also peculiarities that cannot be explained by the assumption that the 
	flow is an illusion caused by the fact that the emission traces the advance of the 
	physical conditions. The ricochets observed in the movie are difficult to explain only 
	by the change in conditions, especially since the neighbouring features are exhibiting 
	kinematics in line with the global behaviour.

	There are features though where the effects of the change in physical conditions are 
	apparent in their behaviour. Usually this is demonstrated with changes of the 
	intensity within the masering spots. As the spots move following the global motion, 
	the intensity peaks are redistributed within them in such a way that they resemble 
	the changes in intensity in the lights of a christmas tree, thus the name `christmas 
	tree effect'. This is observed in many of the filaments in the western part of the 
	ring from $\phi$$\sim$1.56 to the end of the movie.

	\subsection{Comparison with Models}		
	We will compare our observations with the latest and more complete model currently 
	available, that of \cite{gray_2}. Results from TX Cam \citep{gonidakis_1} and R Cas 
	\citep{assaf_1} have been previously compared with the same model. There are 
	significant deviations between the model and our observational data. The model 
	predicts that the diameter of the masering ring should be about twice the size of the 
	stellar photosphere; however, there are variations around this value according to 
	observations \citep{cotton_1, wittkowski_1}. If we assume a stellar photosphere of 
	4.33 mas for TX Cam (as derived from \cite{cahn_1} for the distance of 390 pc adopted 
	here) it appears that the masering shell is located at distances greater than 
	2$R_{\star}$. The maximum distance of the inner shell is 16.24 mas (6.33 au or 
	3.74$R_{\star}$, the minimum is 10.03 mas (3.91 au or 2.32$R_{\star}$) while the 
	weighted mean value over all our observations is only 13.27 mas (5.17 mas or 
	3.07$R_{\star}$). 

	An inconsistency between the model and our observations, is the phase at which the 
	ring starts contracting. This structural change, according to \cite{gray_2}, happens 
	between phases 0.1 and 0.25, when the ring starts decreasing in radius but increases 
	in intensity. In our data, such a structural change should be demonstrated as a 
	decrease in the inner shell boundary radius. However, contraction of the inner shell 
	boundary is not observed at the phases predicted by \cite{gray_2}. In 
	contradistinction, the behaviour of the outer boundary is in line with this result; 
	the outer shell boundary radius starts decreasing at $\phi\sim$ 1.1, 2 and 3.1 
	respectively for each cycle. During the same phases (\phase\similar0.1 to 0.25), the 
	model predicts an increase in the spectral output and this is observed in all the 
	cycles covered by our data. The model also predicts a decrease in intensity between 
	$\phi\sim$~0.25-0.4 - this is also observed in our campaign. Another property that is 
	not apparent in our data during these phases is the shift from a large number of 
	spots of modest brightness to a smaller amount of spots, some of which appear very 
	bright in the model. 

	There are a number of assumptions made in the creation of the model that could cause 
	the disagreement with our data. The model assumes spherical symmetry for both the star 
	and the shock, something that might not be true for TX Cam since, as described in 
	\S~\ref{sec:bipolarity}, there is strong evidence of bipolar outflow. As a matter of 
	fact, asymmetries in Miras might be quite common and there is an increasing amount 
	of observational evidence supporting this. CO(2-1) observations by \cite{josselin_1} 
	in $o$ Ceti revealed strong asymmetries in the gas distribution that could be 
	explained by a bipolar outflow disrupting the spherical envelope. \cite{thompson_1} 
	observed R Tri with the Palomar Testbed Interferometer and found that their data are 
	not in agreement with a spherically symmetric model. \cite{ragland_1} found asymmetric 
	brightness distribution in 16 out of the 56 AGB stars they observed, using the 
	Infrared Optical Telescope Array. Additionally, a restriction only to their 
	well-resolved targets yielded asymmetry detections in 75\% of the AGB and in 100\% of 
	the O-rich stars, allowing the authors to hypothesise that asymmetries might be 
	detectable in all Miras. 

	The model assumes that the abundance of SiO is constant, independent of radius and 
	phase, does not include SiO isotopologues and does not take into account the pulsation 
	of the star. From our data, the distribution of the masers within the zone that could 
	potentially host the effect, reveals that the conditions along the ring are not 
	uniform. The many small gaps observed between individual maser spots could be caused 
	by lower SiO density or lack of velocity coherence. On the other hand, the bigger and 
	locally constant gaps can be attributed to the proposed bipolarity of TX Cam. The 
	model can produce the former; the latter though, when simulated, is just the result 
	of the random selection of the spots since individual spots in the model do not 
	evolve, but the maser positions are rerandomised between each two phases. Moreover, 
	our results for the outer bandary, which compares better with this model to its inner 
	counterpart, suggests that the pulsation of the star influences the borders of the 
	masering zone.
		
\section{Conclusions}
We presented the final version of the movie of the \emission SiO masers
in the extended atmosphere of the Mira Variable TX Cam. The movie covers three complete 
stellar pulsations from phase $\phi$=0.68 to 3.74. This paper is the first in a series 
presenting the results from this campaign and deals with the study of the structure and 
the global kinematic behaviour of the masering zone as well as the existence and 
contribution of shock waves to the observed properties. We conclude the following:\\
(i) The emission is confined to a narrow structure in accordance to previous results.
This structure can change from a ring to an ellipsoid with significant asymmetries.
The inner boundary of the ring reveals the global kinematic behaviour of the gas.
Expansion and contraction are observed, although for the first cycle only expansion is 
observed. \\
(ii) The distribution of the masers is random along the perimeter of the ring. There are 
features that clearly deviate from the overall ordered kinematic behaviour and usually can 
be identified in the movie as splits or ricochets.\\
(iii) Most of the structural and physical properties follow the pulsation of the star, 
revealing a kinematic dependance with phase. Both the width of the ring and the outer 
boundary are correlated with the pulsation phase and the intensity appears to follow the 
stellar pulsation but with a lag.\\
(iv) There are a number of features with kinematics that can be attributed to the 
existence of shock waves. A shock with a velocity of 7 \kms can explain most of the 
observational peculiarities of these features. The morphology and evolution of the outer 
boundary suggests that at these distances shocks are damped.\\
(v) There is strong evidence that the outflow is bipolar. There is a clear difference in 
the expansion velocities along two perpendicular axes. In the SE-NW direction maser spots 
are brighter and are moving faster, while in the NE-SW direction the ring has clear gaps, 
with mainly diffuse emission that moves much slower. \\
(vi) Projection effects contribute to the overall appearance of the shell structure as 
revealed by their shifted radial velocities.\\
(vii) The agreement of our results with the current model seems to be dependent on our 
definition of the ring radius. The outer boundary seems to be following the behaviour 
predicted by the model of \cite{gray_2} better than the inner shell. 
   
\vspace{1cm}

We acknowledge with thanks data from the AAVSO International Database
based on observations submitted to AAVSO by variable star observers
worldwide.

\begin{appendix}
\section{Total Intensity Frames}
\begin{figure*}
\centering
\includegraphics[trim=3mm 90mm 0mm 0mm, clip, scale=0.90]{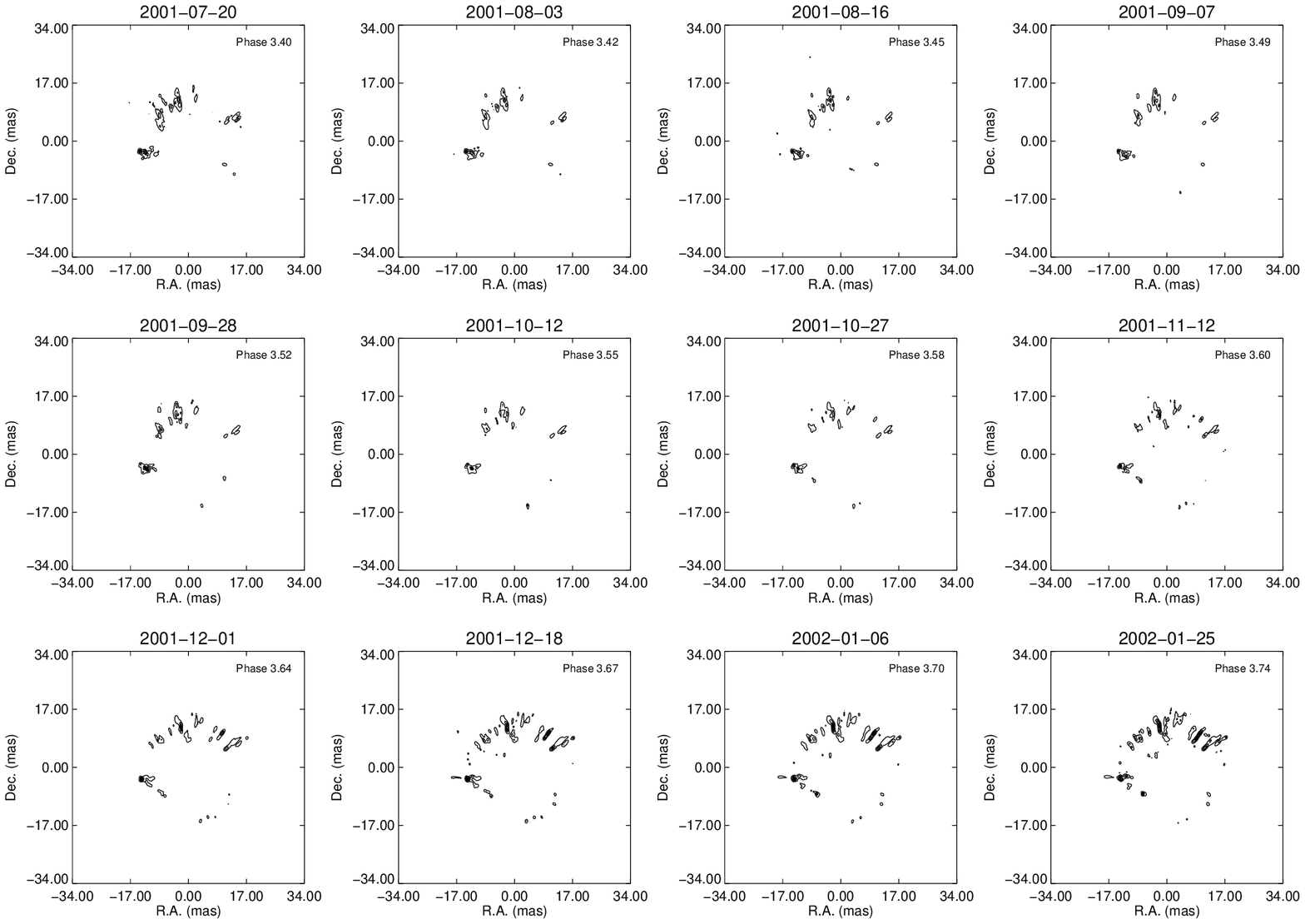}
\caption{As explained in Appendix A}
\label{fig:fig_6}
\end{figure*}

Fig.~\ref{fig:fig_6} is a collage of the total intensity frames. The phase of each frame 
is written at the top right of each contour plot and the date that the data were taken, 
above each frame. The lower contour value is 0.7 Jy/beam and the step between each contour 
is 2.22 Jy/beam until a maximum value of 45.5484 Jy/beam. For a better viewing result, the 
square root of the total intensity maps were used in the compilation of the movie, in 
order to reduce the dynamic range. However, the contour plots presented in this figure 
are the total intensity maps. The six features discussed in the text are also marked here. 
They are enclosed in rectangles at the epoch when they firstly appear or when a change in their 
kinematic behaviour is firstly clearly observed. The number next to the rectangle is the 
number of the feature as used in the text.

\begin{figure*}
\centering
\includegraphics[trim=3mm 0mm 0mm 0mm, clip, scale=0.90]{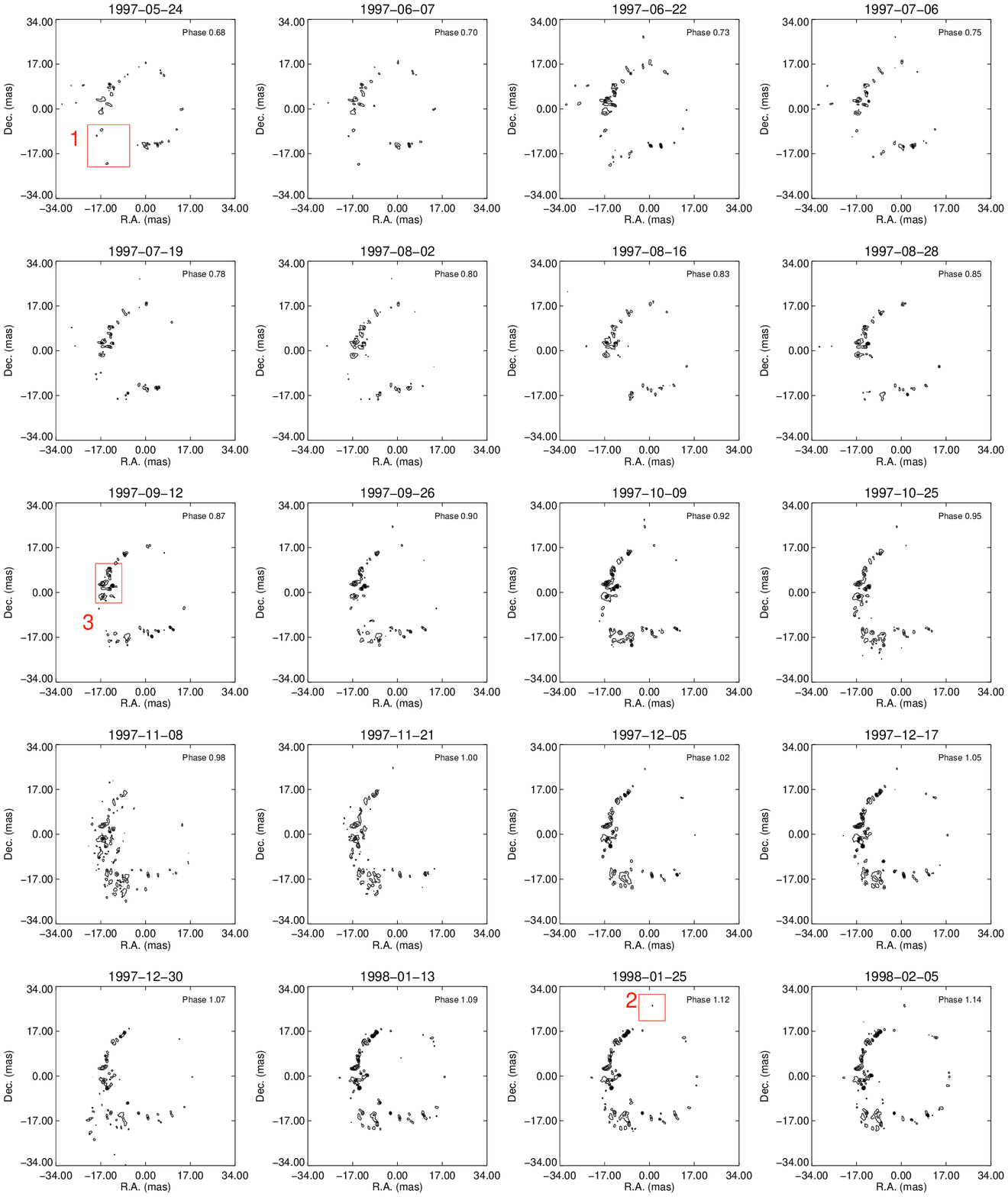}\\
\end{figure*}
\begin{figure*}
\centering
\includegraphics[trim=3mm 0mm 0mm 0mm, clip, scale=0.90]{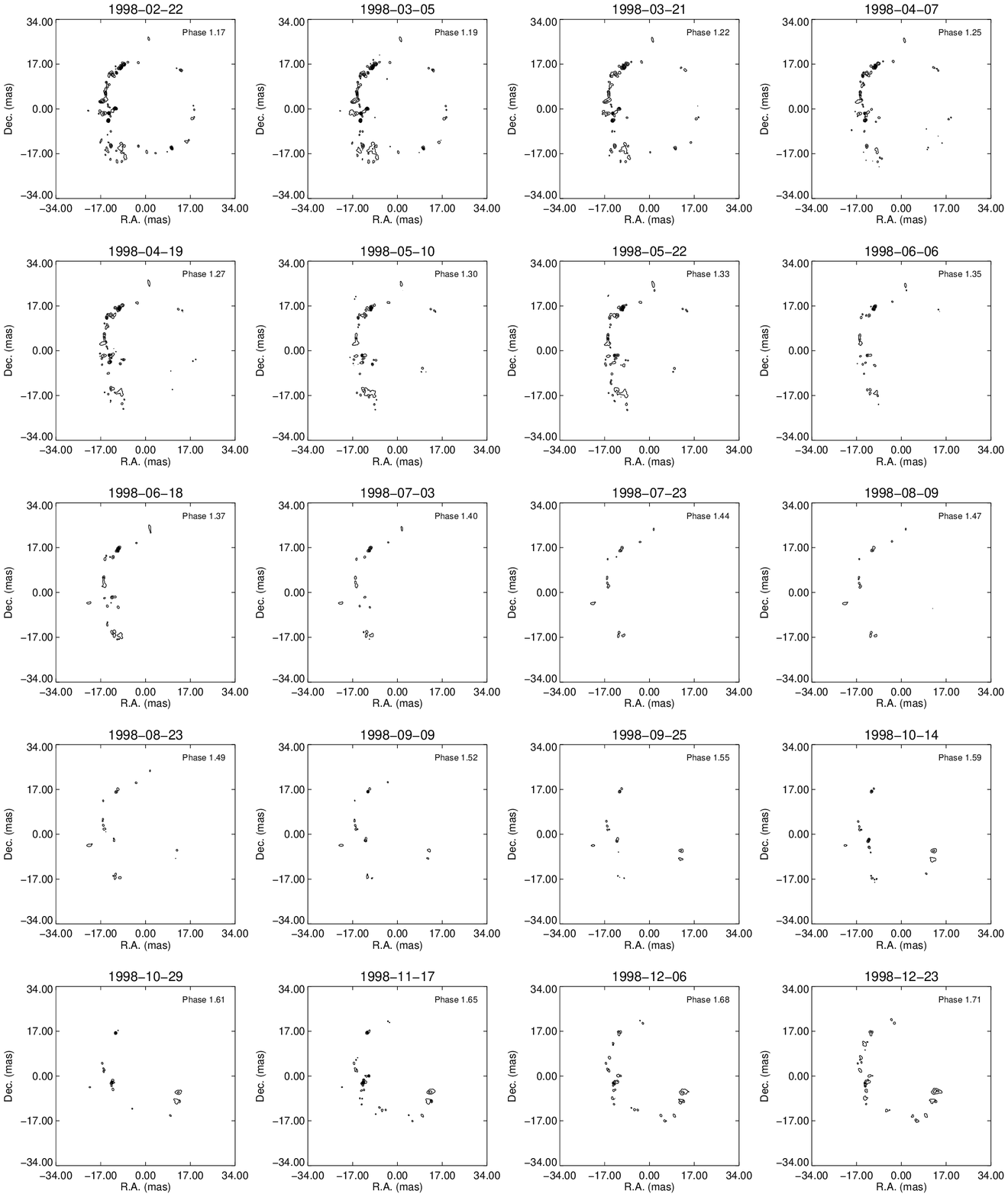}
\end{figure*}
\begin{figure*}
\centering
\includegraphics[trim=3mm 0mm 0mm 0mm, clip, scale=0.90]{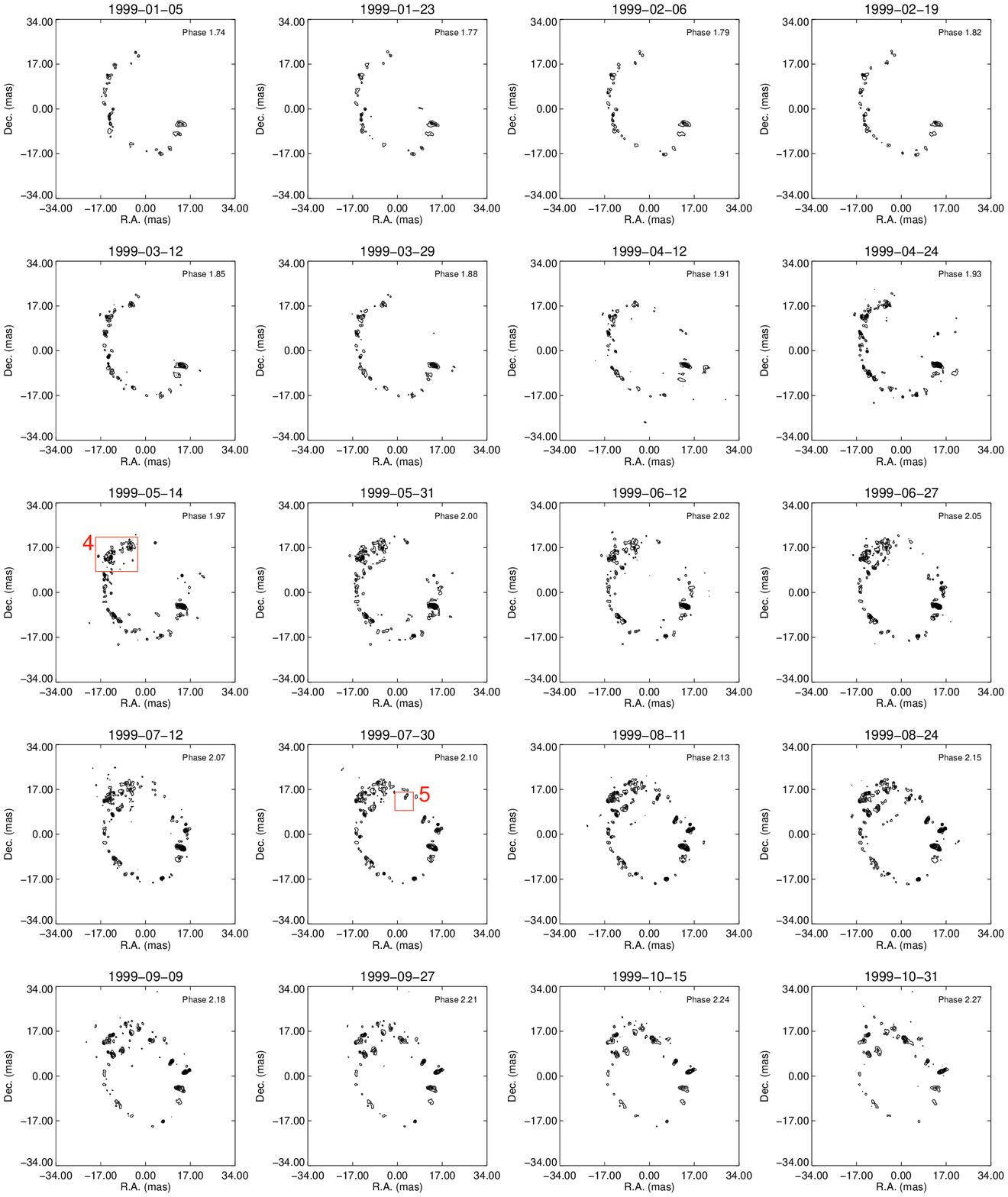}
\end{figure*}		
\begin{figure*}
\centering
\end{figure*}
\begin{figure*}
\centering
\includegraphics[trim=3mm 0mm 0mm 0mm, clip, scale=0.90]{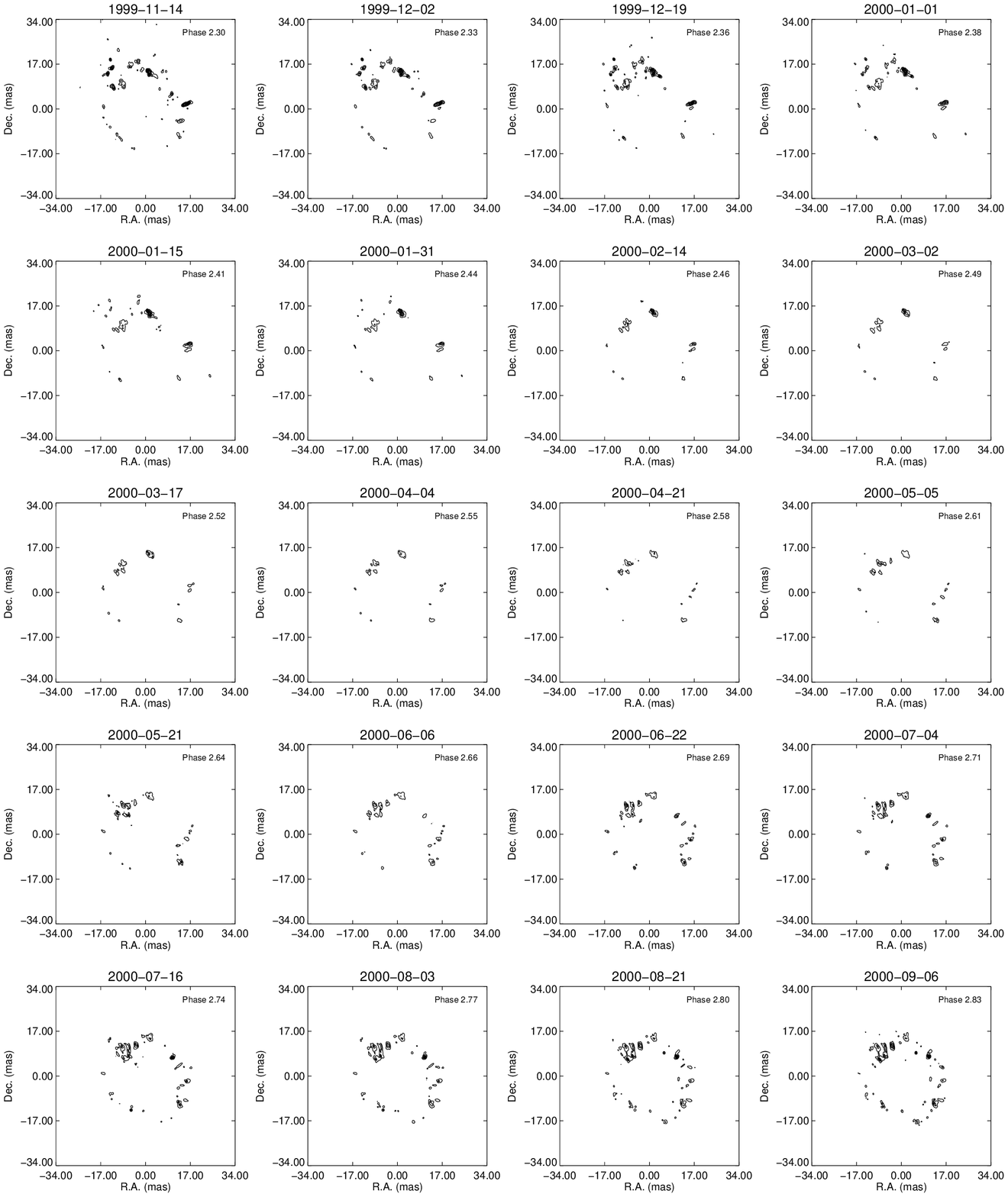}
\end{figure*}
\begin{figure*}
\centering
\includegraphics[trim=3mm 0mm 0mm 0mm, clip, scale=0.90]{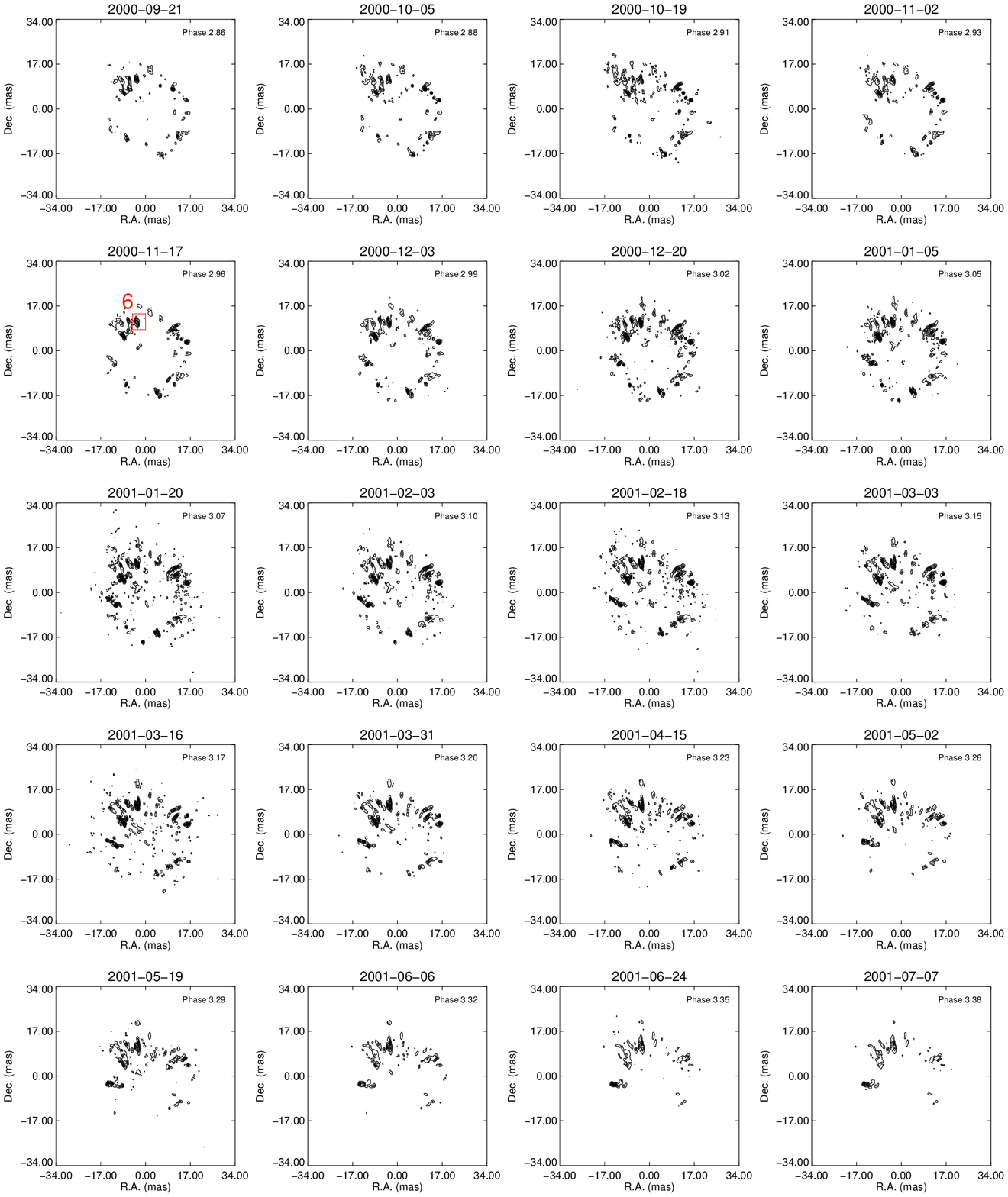}
\end{figure*}
		
\end{appendix}

\end{document}